\documentclass[pra,aps,twocolumn,superscriptaddress,longbibliography]{revtex4-1}
\usepackage[colorlinks=true, citecolor=red, urlcolor=blue ]{hyperref}
\usepackage{graphicx}
\usepackage{bm}
\usepackage{amsmath,amsfonts}
\usepackage{amsthm}
\usepackage{hyperref}
\usepackage{xcolor}
\hypersetup{
    urlcolor=magenta,
    citecolor=blue
           }

\usepackage{braket}
\theoremstyle{plain}
\usepackage{color}
\usepackage{amssymb}
\usepackage{amsthm}
\usepackage{amsfonts}
\usepackage{float}
\usepackage{tabularx}
\usepackage{graphicx}
\usepackage[export]{adjustbox}

\usepackage{mathtools}
\usepackage{esvect}
\usepackage{wrapfig}
\usepackage{amsthm}
\usepackage{verbatim}
\usepackage{bbm}
\usepackage[normalem]{ulem}

\usepackage{enumitem}
\usepackage{fmtcount}
\usepackage{booktabs}
\usepackage{csquotes}
\usepackage{epsfig}

\usepackage{tabularx}
\usepackage{graphicx}
\usepackage{amsmath}
\usepackage{braket}
\usepackage{latexsym}
\usepackage{bm}
\usepackage{graphics,epstopdf}
\usepackage{enumitem}
\usepackage{fmtcount}
\usepackage{booktabs}
\usepackage{csquotes}
\usepackage{epsfig}

\theoremstyle{plain}

\def\bea{\begin{eqnarray}}
\def\eea{\end{eqnarray}}
\def\ba{\begin{array}}
\def\ea{\end{array}}
\def\n{\nonumber}

\def\beq{\begin{equation}}
\def\eeq{\end{equation}}

\usepackage[normalem]{ulem}
\usepackage{float}
\usepackage{graphicx}  
\usepackage{dcolumn}          
\usepackage{amssymb}
\usepackage{appendix}
\usepackage{physics}   
\usepackage{mathtools}
\usepackage{esvect}
\usepackage{wrapfig}
\usepackage{amsthm}
\usepackage{verbatim}
\usepackage{bbm}

\usepackage[mathscr]{euscript}
\def\Tr{\operatorname{Tr}}

\def\({\left(}
\def\){\right)}
\def\[{\left[}
\def\]{\right]}

\begin{document}
%
\title{Approximate second laws and  energy extraction from quantum batteries}
\author{Debarupa Saha}
\affiliation{Harish-Chandra Research Institute,  A CI of Homi Bhabha National Institute, Chhatnag Road, Jhunsi, Prayagraj  211019, India}
\author{Aparajita Bhattacharyya}
\affiliation{Harish-Chandra Research Institute,  A CI of Homi Bhabha National Institute, Chhatnag Road, Jhunsi, Prayagraj  211019, India}
\author{Ujjwal Sen}
\affiliation{Harish-Chandra Research Institute,  A CI of Homi Bhabha National Institute, Chhatnag Road, Jhunsi, Prayagraj  211019, India}

\begin{abstract}
Conservation of energy under thermal operations, \textbf{TO}, is ensured by  commutation of the unitary generating such operations with the total Hamiltonian. However in realistic scenarios, perturbations or disturbances in the system are unavoidable, which in turn may alter the commutation relation and hence in succession may affect the physical processes governed by \textbf{TO}. We call the altered set of operations as approximate thermal operations, \textbf{TO}$_\epsilon$, where $\epsilon$ denotes a degree of disturbance. We provide state transformation conditions under such operations, providing what can be referred to as approximate second laws. We show that in presence of feeble perturbations in the system's Hamiltonian, the states transform in such a way that diagonal elements of the system states start talking not only with each other but also with the off-diagonal elements. In parallel, the off-diagonal elements transform in a way such that they start connecting with diagonal elements and other off-diagonal elements. Such cross-talk is disallowed in the unperturbed second laws. As an application, we show that approximate thermal operations may lead to finite ergotropy extraction from quantum batteries, something that the exact ones are unable to.

\end{abstract}
\maketitle
\section{Introduction}
The resource theory of thermodynamics~\cite{RT,RhT,RhT2,RhT3}, studies the laws of state transformation in presence of a thermal bath under the action of a special kind of operations called thermal operations, \textbf{TO}~\cite{TO,TO2,TO3,TO4}. These are the set of free operations relevant to the scenario, and cannot generate the resource, which here is ``athermality''. Applications of thermodynamic resource theory may be found in works like~\cite{ATO1,ATO2}. The class of unitaries that engender $\textbf{TO}$ conserves the total energy of the system and the bath, and also ensures preservation of the system's Gibbs state. Conservation of energy 
requires
commutation of this special class of unitaries with the total Hamiltonian of the system and the bath. For time-dependent Hamiltonians, it is often the average energy that is considered to be conserved~\cite{Avg}.
 Violation of the energy condition in presence of a homogeneous reservoir was analyzed in~\cite{AlTO}. 
 
State transformation under \textbf{TO} is a well studied realm. Ref.~\cite{RT2} provided a necessary and sufficient condition, and called it thermo-majorization for state transformations, whenever the initial and final states are both diagonal in the energy basis. 
 Ref.~\cite{RC} found a family of free energies that must decrease under $\textbf{TO}$, if 
 catalysis is employed for the desired transformation. Ref.~\cite{RTO} derived how the diagonal and the off-diagonal elements of the system state $\rho$, expressed 
 in the energy eigenbasis, transforms under $\textbf{TO}$. 

 
 Quantum batteries~\cite{Er1,QR2,Col} form an emerging field of research, and is potentially important both fundamentally as well as from a technological viewpoint. 
 The basic idea behind a quantum battery is to charge~\cite{Ch1,Ch2,Ch3,Ch4,Ch5,Ch6} or extract energy~\cite{E1,P1,P2,P3,P4,P5,P6,P7} from a  
 quantum system by means of unitary or more general (quantum) operations. Systems that have been proposed or used as quantum batteries include semiconductor quantum dots~\cite{qd}, spins in magnetic fields~\cite{ma}, electronic states of an organic molecule~\cite{OM,OM2}, superconducting qubits~\cite{Scon}, and states of an electromagnetic field in a high-quality photonic cavity~\cite{Pc}. Further experimental progress regarding implementation of quantum batteries can be found in e.g.~\cite{Ex1,Ex2,Ex3}.
 The maximum amount of energy that can be extracted from such systems by means of a given set of operations is termed as ergotropy~\cite{Er1,Er2,Er3} for that set. The states from which no further energy extraction for the given set of operations is possible are called passive states for that set~\cite{P1,P2,P3,P4,P5,P6,P7}.
 
We have two aims in this work. 
The first is to deal with the problem of state transformation under an altered set of thermal operations, and the second is to show that the altered set has implications for energy extraction from quantum batteries. 
We delve into situations where there is modest violation of the energy conservation condition. This is keeping in mind that in realistic scenarios, disruption in the system is unavoidable. This disruption may manifest itself in terms of an extra potential being added into the system's Hamiltonian. This extra potential may not necessarily commute with the global unitaries corresponding to the $\textbf{TO}$, and hence the resultant operation on the system is no longer a $\textbf{TO}$. If the perturbation is very weak, and quantified by a energy-altering term proportional to a real number 
$\epsilon$,  we call such operations as approximate thermal 
 operations, \textbf{TO}$_\epsilon$.

  We derive state transformation conditions under \textbf{TO}$_\epsilon$, which we call approximate second laws.  We show that under 
 \textbf{TO}$_\epsilon$, the state of the system transforms in a way in which the diagonal elements of the state start talking with other diagonal as well as off-diagonal elements. And we also find that the off-diagonal elements transform in a way in which  they start talking to other diagonal and off-diagonal elements, indicating the possibility of quantum coherence~\cite{Co1,Co2,Co3,Co4,Co5}generation.
 
 We find that approximate thermal operations can be put to use 
 within context of quantum batteries.  For a non-degenerate energy spectrum, if we restrict the energy extraction operation to be a \textbf{TO}, then no energy can be extracted irrespective of the choice of the initial state. We 
 find that 
 irrespective of the choice of initial state of the system, a weak perturbation in the system's Hamiltonian may lead to finite energy extraction. 
 
The rest of the paper is organized as follows.
In Sec.~\ref{1}, we put together some notations that will frequently be used later in the paper.
In Sec.~\ref{2}, we describe a bit about \textbf{TO}. Sec.~\ref{3} contains a description of approximate thermal operations. State transformation conditions under thermal operations are discussed in Sec.~\ref{s4}. In Sec.~\ref{5}, we derive the ``approximate second laws''. In Sec.~\ref{6}, we provide an application of approximate thermal operations within the area of quantum batteries. We present a conclusion in Sec.~\ref{lal-jama-gae}.
\section{Notations}
\label{1}
In this section we introduce the notations of the various physical quantities and physical operations used in this paper.
Following are the relevant quantities and their respective notations.
\begin{itemize}
\item Density operator of the system is denoted as $\rho_S$.
\item Density operator of the reservoir or bath is denoted as, $\rho_{B}$.
\item The quantities, $\tau_{B}$ and $\tau_{S}$, represent the thermal states of the bath and the system respectively.
\item The Hamiltonian of the system is denoted by $H_S$ and that of the reservoir by $H_B$. The total Hamiltonian comprising the system and bath is denoted by $H_T$.
\item The Hamiltonian of the perturbed system is denoted as $H'_S$. The total Hamiltonian comprising of the the perturbed system's Hamiltonian and the bath Hamiltonian $H_B$ is denoted as
$H'_T$.
\item We denote the thermal operations as $\Phi^{TO}$ and the corresponding $\epsilon$ modulated thermal operations as $\Phi_{\epsilon}^{TO}$.
\item Ergotropy extracted from the battery using \textbf{TO} is termed as $R_{TO}$ whereas ergotropy extracted by \textbf{TO}$_\epsilon$ is denoted as $R_\epsilon^{TO}$.
\end{itemize} 
\section{Thermal operations}
\label{2}
We present here  a brief description of thermal operation, \textbf{(TO)}~\cite{TO}.
Suppose, there is a system, $S$, with a Hamiltonian, $H_S$, defined in the $d_1$ dimensional Hilbert space, $\mathcal{H}_S$. Initial state of the system is, $\rho_{S}$. There is also a bath or a reservoir, $R$, with Hamiltonian, $H_B$ defined in the $d_2$ dimensional Hilbert space $\mathcal{H}_B$. Initially, the system and the bath is uncorrelated. The bath is at thermal equilibrium with inverse temperature $\beta$. Thus the initial state of the bath is a thermal state given as $\tau_B= \frac{e^{-\beta H_B}}{Z_B}$, where $Z_B$ is the partition function given by $Z_B=\Tr[e^{-\beta H_B}]$.
The system is made to interact with the bath under the action of a global unitary operator $U_{SB}$. Hence the correlated state of the system and the bath at a later time becomes
\begin{equation}
    \rho_{sb}= U_{SB} (\rho_{S}\otimes \tau_{B}) U_{SB}^\dagger.
    \nonumber
\end{equation}
Final state of the system  after evolution therefore becomes
\begin{equation}
    \Phi(\rho_{s})= \Tr_B[U_{sb} (\rho_{S}\otimes \tau_{B}) U_{sb}^\dagger].
    \nonumber
\end{equation}
The notation, $\Tr_B$, indicates the bath has been traced out.
The resultant operation on the system, $\Phi(\cdot)$, is a completely positive trace preserving, (CPTP), operation. In particular such operations will be called \textbf{TO}, denoted by
 $\Phi^{TO}(\cdot)$, if the following conditions are satisfied.
 \begin{enumerate}
     \item The global unitaries, $U_{SB}=U_{SB}^{TO}$, generating such operations must commute with the the total Hamiltonian, $H_T$, i.e.
     \begin{equation}
         [U_{SB}^{TO},H_T]=0,
         \label{C1}
     \end{equation}
     where $H_T=H_S\otimes \mathcal{I}+\mathcal{I}\otimes H_B$. $\mathcal{I}$ is the Identity operator. 
     The above commutation relation ensures that the total energy of the system and bath remains conserved under thermal operations.
     \item The operation must be Gibbs state preserving i.e. if the initial state of the system is a Gibbs state, $ \tau_S=\frac{e^{-\beta H_S}}{Z_S}$, with $Z_S$ being the partition function given by $Z_S=\Tr[e^{-\beta H_S}]$, then $\Phi^{TO}(\tau_S)=\tau_S$.
 \end{enumerate}
These above two conditions dictate thermal operations. In realistic scenario, noise or perturbation in systems are unavoidable, which may result in to violation of these conditions and eventually may affect the physical processes governed by thermal operation.
We particularly focus on state transformation under \textbf{TO} and ergotropy extraction by thermal operation. In the following section we define approximate thermal operations, $\textbf{TO}_\epsilon$ which arises as a consequence of perturbation in system's Hamiltonian.
\section{Approximate thermal operations}
\label{3}
Suppose the Hamiltonian of the system is weakly perturbed. $H'$ is the Hamiltonian representing a frail disturbance in system's Hamiltonian such that the resultant perturbed Hamiltonian of the system now becomes
\begin{equation}
    H'_{S}=H_S+ \epsilon H'.
    \nonumber
\end{equation}
Here $\epsilon$ is a dimensionless parameter that can range from 0 to 1. For very  weakly perturbed system we have $\epsilon\to0$.
Thus the modified total Hamiltonian of the system and the environment can be written as
\begin{equation}
    H'_T= H'_{S}\otimes\mathcal{I}+\mathcal{I}\otimes H_{B}.
    \nonumber
\end{equation}
Under such circumstances, the class of unitary operations $U_{SB}$ no longer commutes with the modified total Hamiltonian. The commutation relation thus becomes
\begin{equation}
   [U_{SB}^{TO},H_T]=\epsilon [U_{SB}^{TO},H'\otimes \mathcal{I}],
   \label{C2}
\end{equation}
where $[U_{SB}^{TO},H'\otimes \mathcal{I}]\neq 0$. Note that at $\epsilon=0$, the commutation relation reduces to that in~\eqref{C1}.
If $\rho'_{S}$ denotes the initial state of the system written in the energy eigenbasis of $H'_{S}$, then the resultant operation on the system upon the action of the global unitary $U_{SB}$, followed by discarding the reservoir, can be written as 
\begin{equation}
     \Phi'(\rho'_{S})=\Tr_B[U_{SB} (\rho'_{S}\otimes \tau_{B}) U_{SB}^\dagger].
    \nonumber
\end{equation}
Whenever $U_{SB}$ and $H'_{S}$ satisfies the commutation relation~\eqref{C2}, we call the class of operations $ \Phi'(\rho'_{s})$  as approximate thermal operations, $\textbf{TO}_{\epsilon}$, denoted by $\Phi_{\epsilon}^{TO}$. \\
Thus given a group of $H_{S}$, $H'$ and $\epsilon$, for every element, $\Phi_{TO}$, in the set of \textbf{TO}, there exists a corresponding element in set of \textbf{TO}$_\epsilon$, denoted by $\Phi_{TO}^{\epsilon}$.
In the succeeding section we give a brief discussion on state transformation under \textbf{TO}.
\section{State transformation under \textbf{TO}}
\label{s4}
 
 It was shown in~\cite{RTO} that 
 if the system Hamiltonian has a non-degenerate Bohr spectra, i.e., there are no degeneracies in the nonzero differences of energy levels,  any off-diagonal element, $\ketbra{i}{j}$ of a state $\rho$ transforms as 
 \begin{equation}
    \Phi_{TO}(\ketbra{i}{j})=\Lambda_{ij}\ketbra{i}{j},
    \label{ndi}
 \end{equation}
 $\Lambda_{ij}$ is a factor by which the off-diagonal elements are damped during the transition.
 Note that the off-diagonal elements don't mix among themselves or with the diagonal elements.

 On the other hand, the diagonal elements, $\ketbra{i}{i}$, of the state $\rho$  mixes among themselves but not with the off-diagonal elements. The diagonal elements transforms as
 \begin{equation}
 \Phi_{TO}(\ketbra{i}{i})=\sum_j P(i\to j)\ketbra{j}{j},
 \label{di}
 \end{equation}
 $P(i\to j)$ is the transition probability of transferring an element from $i$ to $j$.

As a consequence, if the initial state, $\rho$, is diagonal in the energy eigenbasis, the final state after transition also remains diagonal. Thus \textbf{TO} doesn't generate quantum coherence.

 In the next section we show how frail perturbations in the system affects the state transformation. We derive condition for transformation of both diagonal and off-diagonal elements in such a scenario and call them as approximate second laws.
\section{State transformation under approximate thermal operations}
\label{5}
Here we will explore how perturbation in the system can affect state transformation in presence of thermal bath.
The perturbed Hamiltonian of the system is $H'_S$. Assume that the initial state of the system is written in the energy eigenbasis of $H'_S$. If $\{\ket{i'}\}$ is the set of eigenvectors of $H'_S$ with corresponding set of eigenvalues $\{E'_i\}$, diagonal elements of the system is then given by $\ketbra{i'}{i'}$, multiplied by some factor. The off-diagonal elements of the system can be written as $\ketbra{i'}{j'}$, again multiplied by some factors. 
Considering that the Hamiltonian of the system to be fully non-degenerate, we apply first order non-degenerate perturbation theory to find $\{\ket{i'}\}$ in terms of eigenvectors, $\{\ket{i}\}$, of the unperturbed Hamiltonian, $H_S$. We can expand $\{\ket{i'}\}$ up to linear order in $\epsilon$, since $\epsilon$ is a small quantity. The expansion is given by
\begin{equation}
    \ket{i'}=\ket{i} +\epsilon \sum_{j\neq i} \frac{\bra{j}H'\ket{i}}{E_i-E_j}\ket{j},
    \nonumber
\end{equation}
 $E_x$s are the eigenvalues of $H_S$, $\forall x$. The system is now made to interact with the bath, under the action of the global unitary operator $U_{SB}$ such that, 
 $[U_{SB}, H_T]=0$, but $[U_{SB}, H'_T]=\epsilon[U_{SB}, H']\neq 0$. Thus, unlike what is discussed in Sec.~\ref{s4}  the resultant operation on $\rho'_S$ is $\textbf{TO}_\epsilon$. Let us now examine, how the elements of $\rho'_S$ transform under such a scenario.
\subsubsection{How off-diagonal elements transform}
\label{od}
The bath is initially in an equilibrium state,
\begin{equation}
    \tau_{B}=\sum_{R=1}^{d_2} P_{\mathcal{E}_R} \ketbra{\mathcal{E}_R}{\mathcal{E}_R},
    \nonumber
\end{equation}
 $\{\ket{\mathcal{E}_R}\}$ is the set of eigenvectors of the bath Hamiltonian, $H_B$, and $P_{\mathcal{E}_R}$ is the probability of occupying a particular energy level $\mathcal{E}_R$. Note that, $P_{\mathcal{E}_R}=\frac{e^{-\beta\mathcal{E}_R}}{Z_B}$, with $Z_B$ being the partition function given by $Z_B=\Tr[e^{-\beta H_B}]$. Initially the system, $\rho'_{s}$ and the bath is uncorrelated. Later they were made to evolve together upon the action of a global unitary operator $U_{SB}$. Thus the evolved state of the system can be written as
\begin{eqnarray*}
     \Phi^{\epsilon}_{TO}(\rho'_{S})&=&\Tr_B[U_{SB} (\rho'_{S}\otimes \tau_{B}) U_{SB}^\dagger],\\ &=&\sigma'_{S}.
    \nonumber
\end{eqnarray*}
 It can be shown that the off-diagonal elements, $\ketbra{i'}{j'}$ of the state $\rho'_s$ transform as
 \begin{eqnarray}
    &\Phi^{\epsilon}_{TO}(\ketbra{i'}{j'})=
    \Lambda_{ij}\Big[\ketbra{i'}{j'}- \nonumber \\&\epsilon\Big(\sum_{k'\neq i'} 
    \frac{\bra{k'}H'\ket{i'}}{E'_i-E'_k}\ketbra{k'}{j'}+
    \sum_{l'\neq j'} \frac{\bra{j'}H'\ket{l'}}{E'_j-E'_l}\ketbra{i'}{l'}\Big)\Big]+ \nonumber \\
    &\epsilon\Big(\sum_{l\neq j} \frac{\bra{j}H'\ket{l}}{E_j-E_l}\Lambda_{il}\ketbra{i'}{l'}+\sum_{k\neq i} \frac{\bra{k}H'\ket{i}}{E_i-E_k}\Lambda_{kj}\ketbra{k'}{j'}\Big) \nonumber\\&+ O(\epsilon^2),
    \label{ndie}
      \end{eqnarray}
 $\Lambda_{ab}$ with $a=i,k$ and $b=j,l$ are the damping coefficients defined in equation ~\eqref{ndi}. 
The explicit form of $\Lambda_{ab}$, and the derivation of the above equation is given in Appendix~\ref{A1}.
Note that unlike transformation under \textbf{TO}, off-diagonal elements of the state $\rho'_S$ may mix among themselves as well as with the diagonal elements, $\ketbra{i'}{i'}$ and $\ketbra{j'}{j'}$, under the action $\textbf{TO}_\epsilon$. This guarantees generation of coherence in the energy eigenbasis, i.e. if initially $\rho'_S$ was diagonal in the energy basis with zero coherence. The transformed state $\sigma'_S$ will have a finite amount of coherence. In other words off-diagonal elements of $\sigma'_S$  will be non-zero multiplied by a factor of $\epsilon$. This is true even for very weak disturbance and very thin deviation from \textbf{TO}, i.e in the limit $\epsilon\to 0$. Also note that at $\epsilon=0$, Eq.~\ref{ndie} reduces to Eq.~\ref{ndi}.
\subsubsection{How diagonal elements transform}
\label{Dia}
In this section we will examine how the diagonal elements transform under $\textbf{TO}_\epsilon$. It is shown in Appendix~\ref{A2} that an element $\ketbra{i'}{i'}$ of the initial state $\rho'$ transforms as
\begin{equation}
\begin{split}
    \Phi^{\epsilon}_{TO}(\ketbra{i'}{i'})\\&\hspace{-2cm}=\sum_{j}P(i \to j)\Big(\ketbra{j'}{j'}\\&-\epsilon\Big(\sum_{m\neq j} \frac{1}{E'_j-E'_m}\Big(\bra{m'}H'\ket{j'}\ketbra{j'}{m'}\\&+\bra{j'}H'\ket{m'}\ketbra{m'}{j'}\Big)\Big)\\&+\epsilon\sum_{k\neq i} \frac{1}{E_i-E_k}\Big(\Lambda_{ik}\bra{k'}H'\ket{i'}\ketbra{i'}{k'}\\&+\Lambda_{ki}\bra{i'}H'\ket{k'}\ketbra{k'}{i'}\Big)\\&+ O(\epsilon^2).
    \label{die}
    \end{split}
\end{equation}
The quantity, $P(x \to y)$, denotes the transition probability from an energy level $E_x$ to an energy level $E_y$ and $\Lambda_{xy}$ are the usual damping coefficients.
Unlike the case of $\textbf{TO}$, where the diagonal elements only mix among themselves, under  $\textbf{TO}_{\epsilon}$ operations, the diagonal elements may also mix with the off diagonal elements by an infinitesimal amount. The equations~\ref{ndie} and~\ref{die} so derived, which governs the transformation of the off-diagonal and the diagonal elements of the system, together forms the approximate second laws of state transformation under \textbf{TO}$_\epsilon$ operations.

In the next section we will look at the application of \textbf{TO}$_\epsilon$ in the context of energy extraction from quantum batteries.
\begin{figure*}
\vspace{-4cm}
\includegraphics[scale=0.4]{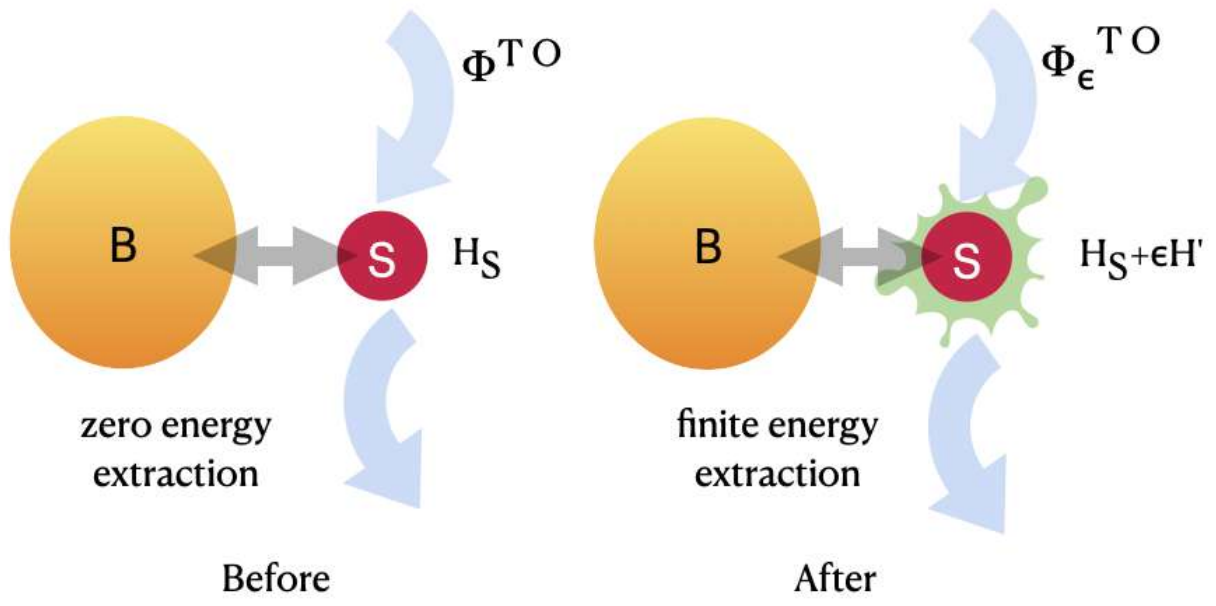}
\vspace{-3cm}
\caption{Two scenarios of energy extraction from a quantum battery. The sub-figure on the left hand side of the schematic depicts the situation when the Hamiltonian of the system, $H_S$, is unperturbed and hence the resultant operation on the system is a $\textbf{TO}$. No energy can be extracted in this case. The sub-figure on the right hand side of the schematic, on the other hand, represents an altered scenario when the Hamiltonian of the system, $H_S$ has been weakly perturbed by a field term $H'$, and hence, the relevant energy extraction operations in this case are the approximate thermal operations. Such operations may lead to a finite energy extraction. }\hspace{-10cm}
\end{figure*}
\section{Application: Quantum batteries}
\label{6}
\subsection{Ergotropy extraction by thermal operation}
Ergotropy refers to the maximum amount of energy that can be extracted from a given quantum system, solely by unitary operation~\cite{Er1,Er2,Er3}.
Mathematically ergotropy, $R(\gamma)$,  of a system $\gamma$ with Hamiltonian $H_J$ can be written as,
\begin{equation}
\begin{split}
    R(\gamma)&=\max_{U}\Big(\Tr[H_J\gamma]-\Tr[H_JU\gamma U^\dagger]\Big)\\&=\Tr[H_J\gamma]-\Tr[H_J\gamma_{p}].
    \end{split}
\end{equation}
where $\gamma_{p}$ is the passive state~\cite{P1,P2,P3,P4,P5,P6,P7}, from which further extraction of energy is not possible by unitary operation.
We restrict the set of unitaries to be the one that satisfies the condition discussed in Sec~\ref{2}, and try to extract energy from the system via \textbf{TO}. In such a case, the maximum energy, $R_{TO}$, that can be extracted using \textbf{TO} is defined as
\begin{equation}
\begin{split}
    R_{TO}(\rho_S)\\&\hspace{-2.2 cm}= \max_{\Phi_{TO}}\Big(\Tr[H_S\rho_S]-\Tr[H_S\Phi_{TO}(\rho_{S})]\Big)\\&\hspace{-2.2cm}=\max_{U_{SB}}\Big(\Tr[H_S\otimes\mathcal{I}\rho_S\otimes \tau_B]\\&-\Tr[H_S\otimes\mathcal{I}U_{SB}\rho_S\otimes \tau_B U^\dagger_{SB}]\Big),
    \end{split}
\end{equation}
where $U_{SB}$ belongs to set of $\textbf{TO}$.
If the total Hamiltonian, $H_{T}$ is non-degenerate, then every eigenvector of it, i.e. $\{\ket{i\mathcal{E}_{R}}\}$, is also a simultaneous eigenvector of $U_{SB}$. Hence one may write 
\begin{equation*}
U_{SB}=\sum_{iR}e^{-I\lambda^R_{i}}\ket{i\mathcal{E}_R}.
\end{equation*}
Thus the maximum extractable energy using $\textbf{TO}$ is given by
\begin{equation}
\begin{split}
     R_{TO}(\rho_S\otimes \tau_B)=
     \max_{\Phi_{TO}}\Big(\Tr[H_S\rho_S]-\Tr[H_S\Phi_{TO}(\rho_{S})]\Big)
\\&\hspace{-8.0cm}=\max_{U_{SB}}(\Tr[H_S\otimes \mathcal{I}(\rho_S\otimes \tau_B)]-\Tr[\rho_S\otimes \tau_B U_{SB}^\dagger H_S\otimes \mathcal{I}U_{SB}])\\&\hspace{-8.0cm}=0.
     \end{split}
     \nonumber
\end{equation}
Thus if the unitary operator employed for energy extraction belongs to set of $\textbf{TO}$, no energy can be extracted. This is true irrespective of the initial state chosen. In the next section we will see that, a mild perturbation in the system Hamiltonian may lead to finite ergotropy extraction by this same class of unitaries.
\subsection{Ergotropy extraction by approximate thermal operation}
\label{toe}
Maximum extractable energy using \textbf{TO}$_\epsilon$ operations, i.e. 
$R_\epsilon^{TO}$, can be written as
\begin{equation}
\begin{split}
  R_\epsilon^{TO}(\rho_S)\\&\hspace{-2.2cm}=\max_{\Phi^\epsilon_{TO}}\Big(\Tr[H'_S\rho_S]-\Tr[H'_S\Phi^\epsilon_{TO}(\rho_{S})]\Big)\\&\hspace{-2.2cm}=\max_{U_{SB}}\Big(\Tr[H'_S\otimes\mathcal{I}\rho_S\otimes \tau_B]\\&-\Tr[H'_S\otimes\mathcal{I}U_{SB}\rho_S\otimes \tau_B U^\dagger_{SB}]\Big).
  \end{split}
\end{equation} 
Any arbitrary state $\rho_S$ can be written in the energy eigenbasis of $H_S$ as $\rho_S=\sum_{m,n}P_{mn}\ketbra{m}{n}$,
and we have $\tau_B=\sum_{K}P_{\mathcal{E}_K}\ketbra{\mathcal{E}_K}{\mathcal{E}_K}$. Decomposing the perturbed Hamiltonian in the energy eigenbasis $\{\ket{i'}\}$ we obtain $H'_S=\sum_{i}h'_{i}\ketbra{i'}{i'}$. It can be shown that 
for a given initial state $\rho_S$, the maximum amount of energy that can be extracted is given by
\begin{equation}
R_\epsilon^{TO}=2\epsilon\sum_{\substack{i,R,j=i+1}} \hspace{-0.3cm}\abs{B^{\mathcal{E}_R}_{ij}} \Big(1+\cos{\theta_{ij}}\Big)\Big]+O(\epsilon^2),
\label{ETO}
\end{equation}
with $B^{\mathcal{E}_R}_{ij}=\bra{j}H'\ket{i}\Big(\frac{h'_{i}}{E_i-E_j}+\frac{h'_{j}}{E_j-E_i}\Big)P_{ij}P_{\mathcal{E}_R}$,  which depends on the initial system and bath parameters, and also on the perturbed potential $H'$. 
We are interested in the limit $\epsilon \to 0$, and therefore neglect the contribution from  the term of the order of $\epsilon^2$. Thus we mainly examine the term of the order of $\epsilon$. Note that this term constitutes a summation over terms which are itself positive, and thus, in general $\abs{B^{\mathcal{E}_R}_{ij}} \left(1+\cos{\theta_{ij}}\right)\geq 0$. It can become zero only when individual terms in the summation are all zero. Since $H_S$ has a non- degenerate eigenspectra, $H'$ does not commute with $H_S$, and hence the factor $\abs{B^{\mathcal{E}_R}_{ij}}\neq 0,  \forall i,j,\mathcal{E}_R$. Thus the only term that can become zero is $(1+\cos{\theta_{ij}})$. Nevertheless, for a given initial state one can always choose a $H'$ such that not every terms $(1+\cos{\theta_{ij}})$ in the summation equals to zero. For example if $H'$ is such that the inner product $\bra{j}H'\ket{i}$ for each $i,j$ is real then $\theta_{ij}=0,  \forall i,j$.
Then the ergotropy is always guaranteed to be greater than zero. Thus there is a possibility of obtaining finite ergotropy, unlike the corresponding scenario of zero energy extraction via \textbf{TO}.  
Also note that at $\epsilon=0$, $ R_{TO}= R^\epsilon_{TO}=0$. The explicit proof of this is provided in Appendix~\ref{A3}.

\section{conclusion}
\label{lal-jama-gae}
We considered
 realistic scenarios where
slight 
disturbances in the system Hamiltonian affects the character of the allowed 
physical processes leading to  state transformations that enhances thermal-bath -induced ergotropy extraction.
We 
emphasized 
that perturbations in the system may 
disrupt
the commutation relation between the unitary, congruous with \textbf{TO},  and the total Hamiltonian. Consequently, the resultant operation on the system will no longer be a thermal operation. We referred such operations as approximate thermal operations, \textbf{TO}$_\epsilon$, where $\epsilon$ denotes the degree of perturbation.

We obtained state transformation conditions both on the diagonal and off-diagonal elements of the system's state under the approximate thermal operations.  We called these conditions as approximate second laws. They are parallel to the state transformation conditions, referred to as second laws, in the case of (exact) thermal operations. The conditions suggest that under approximate thermal operations, the diagonal elements of the system transforms in a way such that they start talking with other diagonal elements as well as with the off-diagonal elements. Alongside, 
the off-diagonal elements transform such that they start talking with other off-diagonal elements as well as the diagonal elements. 
Here, ``$a$ is talking with $b$'' is a colloquial short for the situation when $a$ and $b$ transforms to $a'$ and $b'$ such that $a'$ depends on $b$ (possibly along with $a$). 
Such transformations indicate that approximate thermal operations may lead to quantum coherence generation in the energy eigenbasis, which was 
not possible
in the unperturbed ($\epsilon=0$) case. This is 
possibly an enabling means, 
since quantum coherence is a vital resource in quantum information technologies.
This is especially important, as the 
quantum coherence generation is guaranteed by a  weak and arguably unavoidable disturbance in system's Hamiltonian. 

As an application of the approximate \textbf{TO}, we showed it can 
ensure finite ergotropy extraction from quantum batteries, although its ideal 
cousin (for $\epsilon =0$)
is 
incapable of such energy extraction.


\acknowledgments

We acknowledge discussions with  Ritabrata Sengupta regarding some of the results in the manuscript. 

\appendix
\section{Transformation of off-diagonal elements}
\label{A1}
Here we give the derivation of Eq.~\ref{ndie}. Considering perturbation  theory for non-degenerate eigenspectra, eigenvectors of the perturbed Hamiltonian can be written in terms of the eigenvectors of the unperturbed Hamiltonian as.
\begin{eqnarray}
\label{pt2}
    \ket{i'}&=&\ket{i} +\epsilon \sum_{k\neq i} 
    \frac{\bra{k}H'\ket{i}}{E_i-E_k}\ket{k},\\
    \ket{j'}&=&\ket{j} +\epsilon \sum_{l\neq j} \frac{\bra{l}H'\ket{j}}{E_j-E_l}\ket{l}.
    \label{pt}
\end{eqnarray}
Let the initial state of the system and bath be $\ketbra{i'}{j'}\otimes \tau_B$. After evolution, the element $\ketbra{i'}{j'}$ transforms as 
\begin{equation}
     \Phi^{\epsilon}_{TO}(\ketbra{i'}{j'})=\Tr_B[U_{SB} (\ketbra{i'}{j'}\otimes \tau_{B}) U_{SB}^\dagger].
\end{equation}
Let us first examine the action of $U_{SB}$ on the state $\ketbra{i'}{j'}\otimes \tau_B$. After evolution the total state, $\sigma_{SB}$, can be written as
\begin{equation}
    \sigma_{SB}= U_{SB}\ketbra{i'}{j'}\otimes \tau_{B} U_{SB}^\dagger.
\end{equation}
Substituting the form $\ket{i'}$ and $\ket{j'}$ from Eq.~\ref{pt2} and~\ref{pt} and $\tau_{B}=\sum_{R=1}^{d_2} P_{\mathcal{E}_R} \ketbra{\mathcal{E}_R}{\mathcal{E}_R}$ in  the above equation we get,
\begin{equation}
\begin{split}
    U_{SB}\ketbra{i'}{j'}\otimes \tau_{B} U_{SB}^\dagger &= \sum_{R}P_{\mathcal{E}_R} U_{SB}\Big[\ketbra{i\mathcal{E}_R}{j\mathcal{E}_R}\\ &+\epsilon\Big(\sum_{l\neq j} \frac{\bra{j}H'\ket{l}}{E_j-E_l}\ketbra{i\mathcal{E}_R}{l\mathcal{E}_R}\\ &+\sum_{k\neq i} \frac{\bra{k}H'\ket{i}}{E_i-E_k}\ketbra{k\mathcal{E}_R}{j\mathcal{E}_R}\Big)\Big]U_{SB}^\dagger
    \\ &+O(\epsilon^2).
    \label{n1}
   \end{split}
\end{equation}
Since, $[U_{SB}, H_T]=0$. The entire Hilbert space of the system and the bath, $\mathcal{H}_B\otimes \mathcal{H}_S $, can be decomposed as.
\begin{equation}
   \mathcal{H}_S\otimes \mathcal{H}_B=\bigoplus_E \Bigg(\bigoplus_{E_{S}}\mathcal{H}^{S}_{E_{S}}\otimes\mathcal{H}^{B}_{E-E_{S}}\Bigg).
   \label{Hil}
\end{equation}
$E$ is the total energy and $E_S$ is the energy of the system. See~\cite{RTO} for details. The main idea behind this is that the entire Hilbert space of the system and the environment can be decomposed into blocks of conserved energy. Such that the action of the unitary is just to mix the elements in each block among each other and not with the elements in a different energy block. Keeping this in mind one can write,
\begin{equation}
    U_{SB}\ketbra{i'}{j'}\otimes \tau_{B} U_{SB}^\dagger=\bigoplus_E U_E\ketbra{i'}{j'}\otimes \tau_{B}\bigoplus_{\Bar{E}}U^\dagger_{\Bar{E}}
    \label{Ou}
\end{equation}
$U_E$ and $U_{\Bar{E}}$ acts on blocks of energies $E$ and $\Bar{E}$, respectively, such that.
\begin{equation} U_E\ket{i\mathcal{E}_R}=\sum_{p}\alpha^{pi}_{\mathcal{E}_R}\ket{p}\ket{\mathcal{E}_R+E_i-E_p}\delta_{E,\mathcal{E}_{R}+E_{i}},
\label{it}
\end{equation}
\begin{equation}
\bra{j\mathcal{E}_R}U^\dagger_{\Bar{E}}=\sum_{q}\alpha^{qj*}_{\mathcal{E}_R}\ket{q}\ket{\mathcal{E}_R+E_j-E_q}\delta_{\Bar{E},\mathcal{E}_{R}+E_{j}}.
\nonumber
\end{equation}
We are interested in the reduced density matrix of the system which can be obtained by tracing over the bath. Since trace is linear operation the total reduced density matrix of the system can be written as $\sigma'_{S}=\sum_{x}\sigma^{'x}_{S}$, where $x=1,2,3$ denotes contribution from the first, second and third term of Eq~\ref{n1}. Using the above two equations and Eq.~\ref{Ou} the first term of the Eq.~\ref{n1} turns out to be.
\begin{equation*}  
\begin{split}
\sum_{R}P_{\mathcal{E}_R} U_{SB}\ketbra{i\mathcal{E}_R}{j\mathcal{E}_R}U_{SB}^\dagger=&\\
\sum_{R,p,q}P_{\mathcal{E}_R}\alpha^{pi}_{\mathcal{E}_R}\alpha^{qj*}_{\mathcal{E}_R}\ketbra{p}{q}\otimes\ketbra{\mathcal{E}_R+\omega_{ip}}{\mathcal{E}_R+\omega_{jq}}
\end{split}
\end{equation*}
Such that $E-\Bar{E}=E_i-E_j$. We denote the difference in energy levels, $E_{x}-E_{y}$ as $\omega_{xy}$. Tracing over the bath we get the reduced state of the system to be
\begin{equation}
\sigma^{'1}_{S}=\sum_{p,q|\omega_{ip}=\omega_{jq}}\Big(\sum_{R}P_{\mathcal{E}_R}\alpha^{pi}_{\mathcal{E}_R}\alpha^{qj*}_{\mathcal{E}_R}\Big)\ketbra{p}{q}.
\nonumber
\end{equation}
 Condition over the summation is such that $\omega_{ip}=\omega_{jq}$. For non degenerate Bohr spectra this condition can only be satisfied if $i=p$ and $j=q$, i.e. $\omega_{ip}=\omega_{jq}=0$.
Thus the first term of Eq~\ref{n1} simply gives 
\begin{equation*}
\sigma^{'1}_{S}=\Lambda_{ij}\ketbra{i}{j}.
\end{equation*}
$\Lambda_{ij}=\Big(\sum_{R}P_{\mathcal{E}_R}\alpha^{pi}_{\mathcal{E}_R}\alpha^{qj*}_{\mathcal{E}_R}\Big)$. Next let's look at the second and the third term of Eq.~\ref{n1}. Proceeding in the similar fashion as before. The second term gives.
\begin{equation*}
\hspace{-0.2cm}\sigma^{'2}_{S}=
\epsilon\Big[\sum_{l\neq j} \frac{\bra{j}H'\ket{l}}{E_j-E_l}\sum_{p,m|\omega_{ip}=\omega_{lm}}\Big(\sum_{R}P_{\mathcal{E}_R}\alpha^{pi}_{\mathcal{E}_R}\alpha^{lm*}_{\mathcal{E}_R}\Big)\ketbra{p}{m}\Big]. 
\end{equation*}
Here we have used that, 
\begin{equation*}
\bra{l\mathcal{E}_R}U^\dagger_{\Bar{E}}=\sum_{m}\alpha^{ml*}_{\mathcal{E}_R}\ket{m}\ket{\mathcal{E}_R+E_l-E_m}\delta_{\Bar{E},\mathcal{E}_{R}+E_{j}}. 
\end{equation*}
Again for Bohr non-degenerate eigenspectra, $\omega_{ip}=\omega_{lm}$ is satisfied only when $E_i=E_p$ and  $E_l=E_m$. Putting this we get.
\begin{equation*}
    \sigma^{'2}_{S}=\epsilon\Big[\sum_{l\neq j} \frac{\bra{j}H'\ket{l}}{E_j-E_l}\Lambda_{il}\ketbra{i}{l}\Big].
\end{equation*}
Similarly, from the third term of Eq.~\ref{n1} using,
\begin{equation*} U_E\ket{k\mathcal{E}_R}=\sum_{n}\alpha^{nk}_{\mathcal{E}_R}\ket{n}\ket{\mathcal{E}_R+E_k-E_n}\delta_{E,\mathcal{E}_{R}+E_{k}}. 
\end{equation*}
We get ,
\begin{equation*}
\begin{split}
\sigma^{'3}_{S}=\epsilon\Big[\sum_{k\neq i} \frac{\bra{k}H'\ket{i}}{E_i-E_k}\sum_{n,q|\omega_{kn}=\omega_{jq}}\Big(\sum_{R}P_{\mathcal{E}_R}\alpha^{nk}_{\mathcal{E}_R}\alpha^{qj*}_{\mathcal{E}_R}\Big)\ketbra{n}{q}\Big]. 
\end{split}
\end{equation*}
Again for non-degenerate Bohr spectra this reduces to.
\begin{equation*}
    \sigma^{'3}_{S}=\epsilon\Big[\sum_{k\neq i} \frac{\bra{k}H'\ket{i}}{E_i-E_k}\Lambda_{kj}\ketbra{k}{j}\Big].
\end{equation*}
Making a basis transformation such that the transformed state is written in the eigenbasis of $H'_S$.
We get
\begin{equation*}
\begin{split}
    \sigma'_S=\Phi^{\epsilon}_{TO}(\ketbra{i'}{j'})=
    \Lambda_{ij}\Big[\ketbra{i'}{j'}&-\\\epsilon\Big(\sum_{k'\neq i'} 
    \frac{\bra{k'}H'\ket{i'}}{E'_i-E'_k}\ketbra{k'}{j'}+
    \sum_{l'\neq j'} \frac{\bra{j'}H'\ket{l'}}{E'_j-E'_l}\ketbra{i'}{l'}\Big)\Big]&+ \\
    \epsilon\Big(\sum_{l\neq j} \frac{\bra{j}H'\ket{l}}{E_j-E_l}\Lambda_{il}\ketbra{i'}{l'}+\sum_{k\neq i} \frac{\bra{k}H'\ket{i}}{E_i-E_k}\Lambda_{kj}\ketbra{k'}{j'}\Big)\\&\hspace{-4cm}+ O(\epsilon^2)
      \end{split}
\end{equation*}
Here ends the proof for Eq.~\ref{ndie}.
\section{Transformation of diagonal elements}
\label{A2}
The total evolved state of the system and bath can be written as
\begin{equation}
    \sigma_{SB}= U_{SB}\ketbra{i'}{i'}\otimes \tau_{B} U_{SB}^\dagger
    \nonumber
\end{equation}
Substituting, $\ket{i'}$ from Eq.~\ref{pt2} we get.
\begin{equation*}
\begin{split}
      U_{SB}\ketbra{i'}{i'}\otimes \tau_{B} U_{SB}^\dagger&= \sum_{R}P_{\mathcal{E}_R} U_{SB}\Big[\ketbra{i\mathcal{E}_R}{i\mathcal{E}_R}\\&+
\epsilon\Big(\sum_{k\neq i} \frac{1}{E_i-E_k}\Big(\bra{i}H'\ket{k}\ketbra{i\mathcal{E}_R}{k\mathcal{E}_R}\\&+\bra{k}H'\ket{i}\ketbra{k\mathcal{E}_R}{i\mathcal{E}_R}\Big)\Big)\Big] U^\dagger_{SB}
    \\ &+O(\epsilon^2).
    \end{split}
\end{equation*}
Now using Eq~\ref{Ou} and Eq~\ref{it}, and tracing out the bath, the first term can be written as,
\begin{equation}
\begin{split}
 \Tr_{B} \Big[\sum_{R}P_{\mathcal{E}_R} U_{SB}\Big[\ketbra{i\mathcal{E}_R}{i\mathcal{E}_R}\Big]U^\dagger_{SB}\Big],\\&\hspace{-4.5 cm}=
\sum_{p}\Big(\sum_{R}P_{\mathcal{E}_R}\alpha^{pi}_{\mathcal{E}_R}\alpha^{pi*}_{\mathcal{E}_R}\Big)\ketbra{p}{p}\\&\hspace{-4.5 cm}=
\sum_{p}P(i\to p)\ketbra{p}{p}.
  \end{split}
\end{equation}
 $P(i\to p)=\sum_{R}P_{\mathcal{E}_R}\alpha^{pi}_{\mathcal{E}_R}\alpha^{pi*}_{\mathcal{E}_R}$. Since $p$ is just dummy index we can replace $p$ by $j$ such that $\sum_{j}P(i\to p)\ketbra{p}{p}=\sum_{j}P(i\to j)\ketbra{j}{j}$. Similarly, after tracing the bath the second term can be written as,
\begin{equation}
\begin{split}
 \epsilon \Tr_B\Big[U_{SB}\Big[\sum_{k\neq i} \frac{1}{E_i-E_k}\Big(\bra{i}H'\ket{k}\ketbra{i\mathcal{E}_R}{k\mathcal{E}_R}\\&\hspace{-3cm}+\bra{k}H'\ket{i}\ketbra{k\mathcal{E}_R}{i\mathcal{E}_R}\Big)\Big] U^\dagger_{SB}\Big]\hspace{-4cm}\\&
   \hspace{-5cm}=\epsilon\sum_{k\neq i} \frac{1}{E_i-E_k}\Big(\Lambda_{ik}\bra{k'}H'\ket{i'}\ketbra{i'}{k'}\\&  \hspace{-5cm}+\Lambda_{ki}\bra{i'}H'\ket{k'}\ketbra{k'}{i'}\Big).
    \end{split}
\end{equation}
Making a basis transformation from eigenbasis of $H_{S}$ to eigenbasis of $H'_{S}$, we finally get.
\begin{equation}
\begin{split}
    \Phi^{\epsilon}_{TO}(\ketbra{i'}{i'})\\&\hspace{-2cm}=\sum_{j}P(i \to j)\Big(\ketbra{j'}{j'}\\&-\epsilon\Big(\sum_{m\neq j} \frac{1}{E'_j-E'_m}\Big(\bra{m'}H'\ket{j'}\ketbra{j'}{m'}\\&+\bra{j'}H'\ket{m'}\ketbra{m'}{j'}\Big)\Big)\\&+\epsilon\sum_{k\neq i} \frac{1}{E_i-E_k}\Big(\Lambda_{ik}\bra{k'}H'\ket{i'}\ketbra{i'}{k'}\\&+\Lambda_{ki}\bra{i'}H'\ket{k'}\ketbra{k'}{i'}\Big)\\&+ O(\epsilon^2).
    \end{split}
\end{equation}
This completes the proof of Eq~\ref{die}.
\subsection{Ergotropy extraction by $\Phi^{\epsilon}_{TO}$}
\label{A3}
In this section we will derive the expression for maximum amount of energy extractable under $\epsilon $ modulated \textbf{TO}. 
Beginning with the expression of $ R^\epsilon_{TO}(\rho_S\otimes \tau_B)$ we get.
\begin{equation}
\begin{split}
R^\epsilon_{TO}(\rho_S\otimes \tau_B)\\&\hspace{-2cm}=\max_{U_{SB}}\Big(\Tr[H'_S\otimes\mathcal{I}\rho_S\otimes \tau_B]-\Tr[H'_S\otimes\mathcal{I}U_{SB}\rho_S\otimes \tau_B U^\dagger_{SB}]\Big),\\&\hspace{-2cm}=\max_{U_{SB}}\Big(\Tr[H'_S\otimes\mathcal{I}\rho_S\otimes \tau_B]-\Tr[\rho_S\otimes \tau_B U^\dagger_{SB}H'_S\otimes\mathcal{I}U_{SB}]\Big).
\end{split}
\end{equation}
Let $H'_S=\sum_{i}h'_{i}\ketbra{i'}{i'}$, then 
\begin{equation*}
    H'_S\otimes\mathcal{I}= \sum_{i}h'_i\ketbra{i'\mathcal{E}_R}{i'\mathcal{E}_R}.
\end{equation*}
Putting the expression of $\ket{i'}$ from Eq.~\ref{pt2} we get
\begin{equation*}
\begin{split}
H'_S\otimes\mathcal{I}&=h'_i\Big[\sum_{i}\ketbra{i\mathcal{E}_R}{i\mathcal{E}_R}\\&+
\epsilon\Big(\sum_i\sum_{j\neq i}\frac{1}{E_i-E_j}\Big(\bra{j}H'\ket{i}\ketbra{j\mathcal{E}_R}{i\mathcal{E}_R}\\&+\bra{i}H'\ket{j}\ketbra{i\mathcal{E}_R}{j\mathcal{E}_R}\Big)\Big]\\& +O(\epsilon^2).
\end{split}
\end{equation*}
Putting $U_{SB}=\sum_{iR}e^{-I\lambda^R_{i}}\ket{i\mathcal{E}_R}$ we get,
\begin{equation}
\begin{split}
    U_{SB}H'_S\otimes\mathcal{I} U^\dagger_{SB}=h'_i\Big[\sum_{i}\ketbra{i\mathcal{E}_R}{i\mathcal{E}_R}\hspace{-4.5 cm}\\&+
\epsilon\Big(\sum_i\sum_{j\neq i}\frac{1}{E_i-E_j}\Big(\bra{j}H'\ket{i}\ketbra{j\mathcal{E}_R}{i\mathcal{E}_R}e^{I(\lambda^R_j-\lambda^R_i)}\\&+\bra{i}H'\ket{j}\ketbra{i\mathcal{E}_R}{j\mathcal{E}_R}e^{I(\lambda^R_i-\lambda^R_j)}\Big)\Big]\\& +O(\epsilon^2)
\end{split}
\end{equation}
Note that the diagonal elements of the Hamiltonian remained unchanged under the action of the global unitary. However the off- diagonal elements got multiplied by a phase factor. As a result of which the terms corresponding to the diagonal elements of the Hamiltonian cancels out in the expression of ergotropy, and the resultant expression of ergotropy can now be written as.
\begin{equation}
\begin{split}
R_\epsilon^{TO}&=\epsilon\max_{\{\lambda_i^R\}}\Tr\Big[\\\sum_{i,j\neq i}\frac{1}{E_i-E_j}\Big(\bra{j}H'\ket{i}\ketbra{j\mathcal{E}_R}{i\mathcal{E}_R}(1-e^{I(\lambda^R_j-\lambda^R_i)})\hspace{-8cm}\\&+
\bra{i}H'\ket{j}\ketbra{i\mathcal{E}_R}{j\mathcal{E}_R}(1-e^{I(\lambda^R_j-\lambda^R_i)})\Big)\rho_S\otimes \tau_B \Big]\\&+O(\epsilon^2).
\end{split}
\end{equation}
The maximization is over the set of parameters, $\{\lambda_i^R\}$, with cardinality equals to $d_1*d_2$.
Next we replace $i$ with $j$ and $j$ with $i$ in the second term of the above equation which gives,
\begin{equation}
\begin{split}
    R_\epsilon^{TO}&=\epsilon\max_{\{\lambda_i^R\}}\Tr\Big[\sum_{i,j\neq i}\bra{j}H'\ket{i}\Big(\frac{h'_{i}}{E_i-E_j}\\&+\frac{h'_{j}}{E_j-E_i}\Big)\ketbra{j\mathcal{E}_R}{i\mathcal{E}_R}(1-e^{I(\lambda^R_j-\lambda^R_i)})\rho_S\otimes \tau_B \Big]\\&+O(\epsilon^2).
    \end{split}
\end{equation}
Any arbitrary state $\rho_S$ can be written in the energy basis as $\rho_S=\sum_{\substack{{m,n}}}P_{mn}\ketbra{m}{n}$,
and we have $\tau_B=\sum_{K}P_{\mathcal{E}_K}\ketbra{\mathcal{E}_K}{\mathcal{E}_K}$. Putting the form of $\rho_S$ and $\tau_B$ we get,
\begin{equation}
    \begin{split}
     R_\epsilon^{TO}&=\epsilon\max_{\{\lambda_i^R\}}\Tr\Big[\sum_{\substack{i,j\\n,R|j\neq i}} \hspace{-0.3cm}B^{n\mathcal{E}_R}_{ij} (1-e^{I(\lambda^R_j-\lambda^R_i)})\ketbra{j}{n}\otimes \ketbra{\mathcal{E}_R}{\mathcal{E}_R}\Big]\\&+O(\epsilon^2)
    \end{split}
\end{equation}
 $B^{n\mathcal{E}_R}_{ij}=\bra{j}H'\ket{i}\Big(\frac{h'_{i}}{E_i-E_j}+\frac{h'_{j}}{E_j-E_i}\Big)P_{in}P_{\mathcal{E}_R}$.
Now tracing out the bath and the system we get.
\begin{equation}
    \begin{split}
     R_\epsilon^{TO}&=\epsilon\max_{\{\lambda_i^R\}}\sum_{\substack{i,j\\R|j\neq i}} \hspace{-0.3cm}B^{\mathcal{E}_R}_{ij} (1-e^{I(\lambda^R_j-\lambda^R_i)})\\&+O(\epsilon^2).
    \end{split}
\end{equation}
$B^{\mathcal{E}_R}_{ij}=\bra{j}H'\ket{i}\Big(\frac{h'_{i}}{E_i-E_j}+\frac{h'_{j}}{E_j-E_i}\Big)P_{ij}P_{\mathcal{E}_R}$
Note $B^{\mathcal{E}_R}_{ij}$ is a complex number which can be written as $B^{\mathcal{E}_R}_{ij}=\abs{B^{\mathcal{E}_R}_{ij}}e^{I\theta_{ij}}$.
also note that $\theta_{ij}=-\theta_{ji}$ since $\rho_S$ and $H'$ is Hermitian. Thus the expression of ergotropy simply reduces to 
\begin{equation}
    \begin{split}
     R_\epsilon^{TO}&=\epsilon\max_{\{\lambda_i^R\}}\Big[2\sum_{\substack{i,R,j=i+1}} \hspace{-0.3cm}\abs{B^{\mathcal{E}_R}_{ij}} \Big(\cos{\theta_{ij}}-\cos(\theta_{ij}+\lambda_j^R-\lambda_i^R)\Big)\Big]\\&+O(\epsilon^2).
    \end{split}
\end{equation}
Maximizing over the set of parameters we get
\begin{equation}
R_\epsilon^{TO}=2\epsilon\sum_{\substack{i,R,j=i+1}} \hspace{-0.3cm}\abs{B^{\mathcal{E}_R}_{ij}} \Big(1+\cos{\theta_{ij}}\Big)\Big]+O(\epsilon^2).
\end{equation}
This completes our proof for Eq.~\ref{ETO}.

\vspace{2cm}

\bibliography{ref} 

\begin{thebibliography}{48}%
\makeatletter
\providecommand \@ifxundefined [1]{%
 \@ifx{#1\undefined}
}%
\providecommand \@ifnum [1]{%
 \ifnum #1\expandafter \@firstoftwo
 \else \expandafter \@secondoftwo
 \fi
}%
\providecommand \@ifx [1]{%
 \ifx #1\expandafter \@firstoftwo
 \else \expandafter \@secondoftwo
 \fi
}%
\providecommand \natexlab [1]{#1}%
\providecommand \enquote  [1]{``#1''}%
\providecommand \bibnamefont  [1]{#1}%
\providecommand \bibfnamefont [1]{#1}%
\providecommand \citenamefont [1]{#1}%
\providecommand \href@noop [0]{\@secondoftwo}%
\providecommand \href [0]{\begingroup \@sanitize@url \@href}%
\providecommand \@href[1]{\@@startlink{#1}\@@href}%
\providecommand \@@href[1]{\endgroup#1\@@endlink}%
\providecommand \@sanitize@url [0]{\catcode `\\12\catcode `\$12\catcode `\&12\catcode `\#12\catcode `\^12\catcode `\_12\catcode `\%12\relax}%
\providecommand \@@startlink[1]{}%
\providecommand \@@endlink[0]{}%
\providecommand \url  [0]{\begingroup\@sanitize@url \@url }%
\providecommand \@url [1]{\endgroup\@href {#1}{\urlprefix }}%
\providecommand \urlprefix  [0]{URL }%
\providecommand \Eprint [0]{\href }%
\providecommand \doibase [0]{http://dx.doi.org/}%
\providecommand \selectlanguage [0]{\@gobble}%
\providecommand \bibinfo  [0]{\@secondoftwo}%
\providecommand \bibfield  [0]{\@secondoftwo}%
\providecommand \translation [1]{[#1]}%
\providecommand \BibitemOpen [0]{}%
\providecommand \bibitemStop [0]{}%
\providecommand \bibitemNoStop [0]{.\EOS\space}%
\providecommand \EOS [0]{\spacefactor3000\relax}%
\providecommand \BibitemShut  [1]{\csname bibitem#1\endcsname}%
\let\auto@bib@innerbib\@empty
\bibitem [{\citenamefont {Brand\~ao}\ \emph {et~al.}(2013{\natexlab{a}})\citenamefont {Brand\~ao}, \citenamefont {Horodecki}, \citenamefont {Oppenheim}, \citenamefont {Renes},\ and\ \citenamefont {Spekkens}}]{RT}%
  \BibitemOpen
  \bibfield  {author} {\bibinfo {author} {\bibfnamefont {F.~G. S.~L.}\ \bibnamefont {Brand\~ao}}, \bibinfo {author} {\bibfnamefont {M.}~\bibnamefont {Horodecki}}, \bibinfo {author} {\bibfnamefont {J.}~\bibnamefont {Oppenheim}}, \bibinfo {author} {\bibfnamefont {J.~M.}\ \bibnamefont {Renes}}, \ and\ \bibinfo {author} {\bibfnamefont {R.~W.}\ \bibnamefont {Spekkens}},\ }\bibfield  {title} {\enquote {\bibinfo {title} {Resource theory of quantum states out of thermal equilibrium},}\ }\href {https://link.aps.org/doi/10.1103/PhysRevLett.111.250404} {\bibfield  {journal} {\bibinfo  {journal} {Phys. Rev. Lett.}\ }\textbf {\bibinfo {volume} {111}},\ \bibinfo {pages} {250404} (\bibinfo {year} {2013}{\natexlab{a}})}\BibitemShut {NoStop}%
\bibitem [{\citenamefont {Horodecki}\ and\ \citenamefont {Oppenheim}(2013{\natexlab{a}})}]{RhT}%
  \BibitemOpen
  \bibfield  {author} {\bibinfo {author} {\bibfnamefont {M.}~\bibnamefont {Horodecki}}\ and\ \bibinfo {author} {\bibfnamefont {J.}~\bibnamefont {Oppenheim}},\ }\bibfield  {title} {\enquote {\bibinfo {title} {(quantumness in the context of) resource theories},}\ }\href {https://doi.org/10.1142/S0217979213450197} {\bibfield  {journal} {\bibinfo  {journal} {Int. J. Mod. Phys. B}\ }\textbf {\bibinfo {volume} {27}},\ \bibinfo {pages} {1345019} (\bibinfo {year} {2013}{\natexlab{a}})}\BibitemShut {NoStop}%
\bibitem [{\citenamefont {Ng}\ and\ \citenamefont {Woods}(2019)}]{RhT2}%
  \BibitemOpen
  \bibfield  {author} {\bibinfo {author} {\bibfnamefont {N.~H.~Y.}\ \bibnamefont {Ng}}\ and\ \bibinfo {author} {\bibfnamefont {M.~P.}\ \bibnamefont {Woods}},\ }\bibfield  {title} {\enquote {\bibinfo {title} {Resource theory of quantum thermodynamics: Thermal operations and second laws},}\ }in\ \href {https://doi.org/10.1007/978-3-319-99046-0_26} {\emph {\bibinfo {booktitle} {Thermodynamics in the quantum regime: Fundamental aspects and new directions}}}\ (\bibinfo  {publisher} {Springer},\ \bibinfo {year} {2019})\ p.\ \bibinfo {pages} {650}\BibitemShut {NoStop}%
\bibitem [{\citenamefont {Lostaglio}(2019)}]{RhT3}%
  \BibitemOpen
  \bibfield  {author} {\bibinfo {author} {\bibfnamefont {M.}~\bibnamefont {Lostaglio}},\ }\bibfield  {title} {\enquote {\bibinfo {title} {An introductory review of the resource theory approach to thermodynamics},}\ }\href {https://doi.org/10.1088/1361-6633/ab46e5} {\bibfield  {journal} {\bibinfo  {journal} {Reports on Progress in Physics}\ }\textbf {\bibinfo {volume} {82}},\ \bibinfo {pages} {114001} (\bibinfo {year} {2019})}\BibitemShut {NoStop}%
\bibitem [{\citenamefont {Janzing}\ \emph {et~al.}(2000)\citenamefont {Janzing}, \citenamefont {Wocjan}, \citenamefont {Zeier}, \citenamefont {Geiss},\ and\ \citenamefont {Beth}}]{TO}%
  \BibitemOpen
  \bibfield  {author} {\bibinfo {author} {\bibfnamefont {D.}~\bibnamefont {Janzing}}, \bibinfo {author} {\bibfnamefont {P.}~\bibnamefont {Wocjan}}, \bibinfo {author} {\bibfnamefont {R.}~\bibnamefont {Zeier}}, \bibinfo {author} {\bibfnamefont {R.}~\bibnamefont {Geiss}}, \ and\ \bibinfo {author} {\bibfnamefont {T.}~\bibnamefont {Beth}},\ }\bibfield  {title} {\enquote {\bibinfo {title} {Thermodynamic cost of reliability and low temperatures: Tightening landauer's principle and the second law},}\ }\href {https://doi.org/10.48550/arXiv.quant-ph/0002048} {\bibfield  {journal} {\bibinfo  {journal} {Int. J. Theor. Phys.}\ }\textbf {\bibinfo {volume} {39}},\ \bibinfo {pages} {2717} (\bibinfo {year} {2000})}\BibitemShut {NoStop}%
\bibitem [{\citenamefont {Streater}(2009)}]{TO2}%
  \BibitemOpen
  \bibfield  {author} {\bibinfo {author} {\bibfnamefont {R.~F.}\ \bibnamefont {Streater}},\ }\href@noop {} {\emph {\bibinfo {title} {Statistical dynamics: a stochastic approach to nonequilibrium thermodynamics}}}\ (\bibinfo  {publisher} {World Scientific Publishing Company},\ \bibinfo {year} {2009})\BibitemShut {NoStop}%
\bibitem [{\citenamefont {Horodecki}\ and\ \citenamefont {Oppenheim}(2013{\natexlab{b}})}]{TO3}%
  \BibitemOpen
  \bibfield  {author} {\bibinfo {author} {\bibfnamefont {M.}~\bibnamefont {Horodecki}}\ and\ \bibinfo {author} {\bibfnamefont {J.}~\bibnamefont {Oppenheim}},\ }\bibfield  {title} {\enquote {\bibinfo {title} {Fundamental limitations for quantum and nanoscale thermodynamics},}\ }\href {https://www.nature.com/articles/ncomms3059} {\bibfield  {journal} {\bibinfo  {journal} {Nature communications}\ }\textbf {\bibinfo {volume} {4}},\ \bibinfo {pages} {2059} (\bibinfo {year} {2013}{\natexlab{b}})}\BibitemShut {NoStop}%
\bibitem [{\citenamefont {Brand\~ao}\ \emph {et~al.}(2013{\natexlab{b}})\citenamefont {Brand\~ao}, \citenamefont {Horodecki}, \citenamefont {Oppenheim}, \citenamefont {Renes},\ and\ \citenamefont {Spekkens}}]{TO4}%
  \BibitemOpen
  \bibfield  {author} {\bibinfo {author} {\bibfnamefont {F.~G. S.~L.}\ \bibnamefont {Brand\~ao}}, \bibinfo {author} {\bibfnamefont {M.}~\bibnamefont {Horodecki}}, \bibinfo {author} {\bibfnamefont {J.}~\bibnamefont {Oppenheim}}, \bibinfo {author} {\bibfnamefont {J~M.}\ \bibnamefont {Renes}}, \ and\ \bibinfo {author} {\bibfnamefont {R.~W.}\ \bibnamefont {Spekkens}},\ }\bibfield  {title} {\enquote {\bibinfo {title} {Resource theory of quantum states out of thermal equilibrium},}\ }\href {https://link.aps.org/doi/10.1103/PhysRevLett.111.250404} {\bibfield  {journal} {\bibinfo  {journal} {Phys. Rev. Lett.}\ }\textbf {\bibinfo {volume} {111}},\ \bibinfo {pages} {250404} (\bibinfo {year} {2013}{\natexlab{b}})}\BibitemShut {NoStop}%
\bibitem [{\citenamefont {Yunger~Halpern}\ and\ \citenamefont {Limmer}(2020)}]{ATO1}%
  \BibitemOpen
  \bibfield  {author} {\bibinfo {author} {\bibfnamefont {N.}~\bibnamefont {Yunger~Halpern}}\ and\ \bibinfo {author} {\bibfnamefont {D.~T.}\ \bibnamefont {Limmer}},\ }\bibfield  {title} {\enquote {\bibinfo {title} {Fundamental limitations on photoisomerization from thermodynamic resource theories},}\ }\href {\doibase 10.1103/PhysRevA.101.042116} {\bibfield  {journal} {\bibinfo  {journal} {Phys. Rev. A}\ }\textbf {\bibinfo {volume} {101}},\ \bibinfo {pages} {042116} (\bibinfo {year} {2020})}\BibitemShut {NoStop}%
\bibitem [{\citenamefont {Alhambra}\ \emph {et~al.}(2019{\natexlab{a}})\citenamefont {Alhambra}, \citenamefont {Lostaglio},\ and\ \citenamefont {Perry}}]{ATO2}%
  \BibitemOpen
  \bibfield  {author} {\bibinfo {author} {\bibfnamefont {{\'A}.~M}\ \bibnamefont {Alhambra}}, \bibinfo {author} {\bibfnamefont {M.}~\bibnamefont {Lostaglio}}, \ and\ \bibinfo {author} {\bibfnamefont {C.}~\bibnamefont {Perry}},\ }\bibfield  {title} {\enquote {\bibinfo {title} {Heat-bath algorithmic cooling with optimal thermalization strategies},}\ }\href {https://doi.org/10.22331/q-2019-09-23-188 Focus to learn more} {\bibfield  {journal} {\bibinfo  {journal} {Quantum}\ }\textbf {\bibinfo {volume} {3}},\ \bibinfo {pages} {188} (\bibinfo {year} {2019}{\natexlab{a}})}\BibitemShut {NoStop}%
\bibitem [{\citenamefont {Skrzypczyk}\ \emph {et~al.}(2014)\citenamefont {Skrzypczyk}, \citenamefont {Short},\ and\ \citenamefont {Popescu}}]{Avg}%
  \BibitemOpen
  \bibfield  {author} {\bibinfo {author} {\bibfnamefont {P.}~\bibnamefont {Skrzypczyk}}, \bibinfo {author} {\bibfnamefont {A.~J}\ \bibnamefont {Short}}, \ and\ \bibinfo {author} {\bibfnamefont {S.}~\bibnamefont {Popescu}},\ }\bibfield  {title} {\enquote {\bibinfo {title} {Work extraction and thermodynamics for individual quantum systems},}\ }\href {https://www.nature.com/articles/ncomms5185} {\bibfield  {journal} {\bibinfo  {journal} {Nat. Commun}\ }\textbf {\bibinfo {volume} {5}},\ \bibinfo {pages} {4185} (\bibinfo {year} {2014})}\BibitemShut {NoStop}%
\bibitem [{\citenamefont {Shu}\ \emph {et~al.}(2019)\citenamefont {Shu}, \citenamefont {Cai}, \citenamefont {Seah}, \citenamefont {Nimmrichter},\ and\ \citenamefont {Scarani}}]{AlTO}%
  \BibitemOpen
  \bibfield  {author} {\bibinfo {author} {\bibfnamefont {A.}~\bibnamefont {Shu}}, \bibinfo {author} {\bibfnamefont {Y.}~\bibnamefont {Cai}}, \bibinfo {author} {\bibfnamefont {S.}~\bibnamefont {Seah}}, \bibinfo {author} {\bibfnamefont {S.}~\bibnamefont {Nimmrichter}}, \ and\ \bibinfo {author} {\bibfnamefont {V.}~\bibnamefont {Scarani}},\ }\bibfield  {title} {\enquote {\bibinfo {title} {Almost thermal operations: Inhomogeneous reservoirs},}\ }\href {https://arxiv.org/pdf/1904.08736} {\bibfield  {journal} {\bibinfo  {journal} {Physical Review A}\ }\textbf {\bibinfo {volume} {100}},\ \bibinfo {pages} {042107} (\bibinfo {year} {2019})}\BibitemShut {NoStop}%
\bibitem [{\citenamefont {Horodecki}\ and\ \citenamefont {Oppenheim}(2013{\natexlab{c}})}]{RT2}%
  \BibitemOpen
  \bibfield  {author} {\bibinfo {author} {\bibfnamefont {M.}~\bibnamefont {Horodecki}}\ and\ \bibinfo {author} {\bibfnamefont {J.}~\bibnamefont {Oppenheim}},\ }\bibfield  {title} {\enquote {\bibinfo {title} {Fundamental limitations for quantum and nanoscale thermodynamics},}\ }\href {https://www.nature.com/articles/ncomms3059} {\bibfield  {journal} {\bibinfo  {journal} {Nat. Commun}\ }\textbf {\bibinfo {volume} {4}},\ \bibinfo {pages} {2059} (\bibinfo {year} {2013}{\natexlab{c}})}\BibitemShut {NoStop}%
\bibitem [{\citenamefont {Brandao}\ \emph {et~al.}(2015)\citenamefont {Brandao}, \citenamefont {Horodecki}, \citenamefont {Ng}, \citenamefont {Oppenheim},\ and\ \citenamefont {Wehner}}]{RC}%
  \BibitemOpen
  \bibfield  {author} {\bibinfo {author} {\bibfnamefont {F.}~\bibnamefont {Brandao}}, \bibinfo {author} {\bibfnamefont {M.}~\bibnamefont {Horodecki}}, \bibinfo {author} {\bibfnamefont {N.}~\bibnamefont {Ng}}, \bibinfo {author} {\bibfnamefont {J.}~\bibnamefont {Oppenheim}}, \ and\ \bibinfo {author} {\bibfnamefont {S.}~\bibnamefont {Wehner}},\ }\bibfield  {title} {\enquote {\bibinfo {title} {The second laws of quantum thermodynamics},}\ }\href {https://doi.org/10.1073/pnas.1411728112} {\bibfield  {journal} {\bibinfo  {journal} {PNAS}\ }\textbf {\bibinfo {volume} {112}},\ \bibinfo {pages} {3275} (\bibinfo {year} {2015})}\BibitemShut {NoStop}%
\bibitem [{\citenamefont {\ifmmode \acute{C}\else \'{C}\fi{}wikli\ifmmode~\acute{n}\else \'{n}\fi{}ski}\ \emph {et~al.}(2015)\citenamefont {\ifmmode \acute{C}\else \'{C}\fi{}wikli\ifmmode~\acute{n}\else \'{n}\fi{}ski}, \citenamefont {Studzi\ifmmode~\acute{n}\else \'{n}\fi{}ski}, \citenamefont {Horodecki},\ and\ \citenamefont {Oppenheim}}]{RTO}%
  \BibitemOpen
  \bibfield  {author} {\bibinfo {author} {\bibfnamefont {P.}~\bibnamefont {\ifmmode \acute{C}\else \'{C}\fi{}wikli\ifmmode~\acute{n}\else \'{n}\fi{}ski}}, \bibinfo {author} {\bibfnamefont {M.}~\bibnamefont {Studzi\ifmmode~\acute{n}\else \'{n}\fi{}ski}}, \bibinfo {author} {\bibfnamefont {M.}~\bibnamefont {Horodecki}}, \ and\ \bibinfo {author} {\bibfnamefont {J.}~\bibnamefont {Oppenheim}},\ }\bibfield  {title} {\enquote {\bibinfo {title} {Limitations on the evolution of quantum coherences: Towards fully quantum second laws of thermodynamics},}\ }\href {https://link.aps.org/doi/10.1103/PhysRevLett.115.210403} {\bibfield  {journal} {\bibinfo  {journal} {Phys. Rev. Lett.}\ }\textbf {\bibinfo {volume} {115}},\ \bibinfo {pages} {210403} (\bibinfo {year} {2015})}\BibitemShut {NoStop}%
\bibitem [{\citenamefont {Alicki}\ and\ \citenamefont {Fannes}(2013)}]{Er1}%
  \BibitemOpen
  \bibfield  {author} {\bibinfo {author} {\bibfnamefont {R.}~\bibnamefont {Alicki}}\ and\ \bibinfo {author} {\bibfnamefont {M.}~\bibnamefont {Fannes}},\ }\bibfield  {title} {\enquote {\bibinfo {title} {Entanglement boost for extractable work from ensembles of quantum batteries},}\ }\href {https://link.aps.org/doi/10.1103/PhysRevE.87.042123} {\bibfield  {journal} {\bibinfo  {journal} {Phys. Rev. E}\ }\textbf {\bibinfo {volume} {87}},\ \bibinfo {pages} {042123} (\bibinfo {year} {2013})}\BibitemShut {NoStop}%
\bibitem [{\citenamefont {Bhattacharjee}\ and\ \citenamefont {Dutta}(2021{\natexlab{a}})}]{QR2}%
  \BibitemOpen
  \bibfield  {author} {\bibinfo {author} {\bibfnamefont {S.}~\bibnamefont {Bhattacharjee}}\ and\ \bibinfo {author} {\bibfnamefont {A.}~\bibnamefont {Dutta}},\ }\bibfield  {title} {\enquote {\bibinfo {title} {Quantum thermal machines and batteries},}\ }\href {https://link.springer.com/article/10.1140/epjb/s10051-021-00235-3} {\bibfield  {journal} {\bibinfo  {journal} {EPJ B.}\ }\textbf {\bibinfo {volume} {94}},\ \bibinfo {pages} {42} (\bibinfo {year} {2021}{\natexlab{a}})}\BibitemShut {NoStop}%
\bibitem [{\citenamefont {Campaioli}\ \emph {et~al.}(2024)\citenamefont {Campaioli}, \citenamefont {Gherardini}, \citenamefont {Quach}, \citenamefont {Polini},\ and\ \citenamefont {Andolina}}]{Col}%
  \BibitemOpen
  \bibfield  {author} {\bibinfo {author} {\bibfnamefont {Francesco}\ \bibnamefont {Campaioli}}, \bibinfo {author} {\bibfnamefont {Stefano}\ \bibnamefont {Gherardini}}, \bibinfo {author} {\bibfnamefont {James~Q.}\ \bibnamefont {Quach}}, \bibinfo {author} {\bibfnamefont {Marco}\ \bibnamefont {Polini}}, \ and\ \bibinfo {author} {\bibfnamefont {Gian~Marcello}\ \bibnamefont {Andolina}},\ }\bibfield  {title} {\enquote {\bibinfo {title} {Colloquium: Quantum batteries},}\ }\href {\doibase 10.1103/RevModPhys.96.031001} {\bibfield  {journal} {\bibinfo  {journal} {Rev. Mod. Phys.}\ }\textbf {\bibinfo {volume} {96}},\ \bibinfo {pages} {031001} (\bibinfo {year} {2024})}\BibitemShut {NoStop}%
\bibitem [{\citenamefont {Binder}\ \emph {et~al.}(2015)\citenamefont {Binder}, \citenamefont {Vinjanampathy}, \citenamefont {Modi},\ and\ \citenamefont {Goold}}]{Ch1}%
  \BibitemOpen
  \bibfield  {author} {\bibinfo {author} {\bibfnamefont {F.~C.}\ \bibnamefont {Binder}}, \bibinfo {author} {\bibfnamefont {S.}~\bibnamefont {Vinjanampathy}}, \bibinfo {author} {\bibfnamefont {K.}~\bibnamefont {Modi}}, \ and\ \bibinfo {author} {\bibfnamefont {J.}~\bibnamefont {Goold}},\ }\bibfield  {title} {\enquote {\bibinfo {title} {Quantacell: powerful charging of quantum batteries},}\ }\href {https://iopscience.iop.org/article/10.1088/1367-2630/17/7/075015/meta} {\bibfield  {journal} {\bibinfo  {journal} {NJP}\ }\textbf {\bibinfo {volume} {17}},\ \bibinfo {pages} {075015} (\bibinfo {year} {2015})}\BibitemShut {NoStop}%
\bibitem [{\citenamefont {Campaioli}\ \emph {et~al.}(2017)\citenamefont {Campaioli}, \citenamefont {Pollock}, \citenamefont {Binder}, \citenamefont {C\'eleri}, \citenamefont {Goold}, \citenamefont {Vinjanampathy},\ and\ \citenamefont {Modi}}]{Ch2}%
  \BibitemOpen
  \bibfield  {author} {\bibinfo {author} {\bibfnamefont {F.}~\bibnamefont {Campaioli}}, \bibinfo {author} {\bibfnamefont {F.~A.}\ \bibnamefont {Pollock}}, \bibinfo {author} {\bibfnamefont {F.~C.}\ \bibnamefont {Binder}}, \bibinfo {author} {\bibfnamefont {L.}~\bibnamefont {C\'eleri}}, \bibinfo {author} {\bibfnamefont {John}\ \bibnamefont {Goold}}, \bibinfo {author} {\bibfnamefont {Sai}\ \bibnamefont {Vinjanampathy}}, \ and\ \bibinfo {author} {\bibfnamefont {K.}~\bibnamefont {Modi}},\ }\bibfield  {title} {\enquote {\bibinfo {title} {Enhancing the charging power of quantum batteries},}\ }\href {https://link.aps.org/doi/10.1103/PhysRevLett.118.150601} {\bibfield  {journal} {\bibinfo  {journal} {Phys. Rev. Lett.}\ }\textbf {\bibinfo {volume} {118}},\ \bibinfo {pages} {150601} (\bibinfo {year} {2017})}\BibitemShut {NoStop}%
\bibitem [{\citenamefont {Ferraro}\ \emph {et~al.}(2018)\citenamefont {Ferraro}, \citenamefont {Campisi}, \citenamefont {Andolina}, \citenamefont {Pellegrini},\ and\ \citenamefont {Polini}}]{Ch3}%
  \BibitemOpen
  \bibfield  {author} {\bibinfo {author} {\bibfnamefont {Dario}\ \bibnamefont {Ferraro}}, \bibinfo {author} {\bibfnamefont {Michele}\ \bibnamefont {Campisi}}, \bibinfo {author} {\bibfnamefont {Gian~Marcello}\ \bibnamefont {Andolina}}, \bibinfo {author} {\bibfnamefont {Vittorio}\ \bibnamefont {Pellegrini}}, \ and\ \bibinfo {author} {\bibfnamefont {Marco}\ \bibnamefont {Polini}},\ }\bibfield  {title} {\enquote {\bibinfo {title} {High-power collective charging of a solid-state quantum battery},}\ }\href {\doibase 10.1103/PhysRevLett.120.117702} {\bibfield  {journal} {\bibinfo  {journal} {Phys. Rev. Lett.}\ }\textbf {\bibinfo {volume} {120}},\ \bibinfo {pages} {117702} (\bibinfo {year} {2018})}\BibitemShut {NoStop}%
\bibitem [{\citenamefont {Le}\ \emph {et~al.}(2018{\natexlab{a}})\citenamefont {Le}, \citenamefont {Levinsen}, \citenamefont {Modi}, \citenamefont {Parish},\ and\ \citenamefont {Pollock}}]{Ch4}%
  \BibitemOpen
  \bibfield  {author} {\bibinfo {author} {\bibfnamefont {Thao~P.}\ \bibnamefont {Le}}, \bibinfo {author} {\bibfnamefont {Jesper}\ \bibnamefont {Levinsen}}, \bibinfo {author} {\bibfnamefont {Kavan}\ \bibnamefont {Modi}}, \bibinfo {author} {\bibfnamefont {Meera~M.}\ \bibnamefont {Parish}}, \ and\ \bibinfo {author} {\bibfnamefont {Felix~A.}\ \bibnamefont {Pollock}},\ }\bibfield  {title} {\enquote {\bibinfo {title} {Spin-chain model of a many-body quantum battery},}\ }\href {\doibase 10.1103/PhysRevA.97.022106} {\bibfield  {journal} {\bibinfo  {journal} {Phys. Rev. A}\ }\textbf {\bibinfo {volume} {97}},\ \bibinfo {pages} {022106} (\bibinfo {year} {2018}{\natexlab{a}})}\BibitemShut {NoStop}%
\bibitem [{\citenamefont {Rossini}\ \emph {et~al.}(2019)\citenamefont {Rossini}, \citenamefont {Andolina},\ and\ \citenamefont {Polini}}]{Ch5}%
  \BibitemOpen
  \bibfield  {author} {\bibinfo {author} {\bibfnamefont {D.}~\bibnamefont {Rossini}}, \bibinfo {author} {\bibfnamefont {G.~M.}\ \bibnamefont {Andolina}}, \ and\ \bibinfo {author} {\bibfnamefont {M.}~\bibnamefont {Polini}},\ }\bibfield  {title} {\enquote {\bibinfo {title} {Many-body localized quantum batteries},}\ }\href {https://link.aps.org/doi/10.1103/PhysRevB.100.115142} {\bibfield  {journal} {\bibinfo  {journal} {Phys. Rev. B}\ }\textbf {\bibinfo {volume} {100}},\ \bibinfo {pages} {115142} (\bibinfo {year} {2019})}\BibitemShut {NoStop}%
\bibitem [{\citenamefont {Farina}\ \emph {et~al.}(2019)\citenamefont {Farina}, \citenamefont {Andolina}, \citenamefont {Mari}, \citenamefont {Polini},\ and\ \citenamefont {Giovannetti}}]{Ch6}%
  \BibitemOpen
  \bibfield  {author} {\bibinfo {author} {\bibfnamefont {D.}~\bibnamefont {Farina}}, \bibinfo {author} {\bibfnamefont {G.~M.}\ \bibnamefont {Andolina}}, \bibinfo {author} {\bibfnamefont {A.}~\bibnamefont {Mari}}, \bibinfo {author} {\bibfnamefont {M.}~\bibnamefont {Polini}}, \ and\ \bibinfo {author} {\bibfnamefont {V.}~\bibnamefont {Giovannetti}},\ }\bibfield  {title} {\enquote {\bibinfo {title} {Charger-mediated energy transfer for quantum batteries: An open-system approach},}\ }\href {https://link.aps.org/doi/10.1103/PhysRevB.99.035421} {\bibfield  {journal} {\bibinfo  {journal} {Phys. Rev. B}\ }\textbf {\bibinfo {volume} {99}},\ \bibinfo {pages} {035421} (\bibinfo {year} {2019})}\BibitemShut {NoStop}%
\bibitem [{\citenamefont {Allahverdyan}\ \emph {et~al.}(2004{\natexlab{a}})\citenamefont {Allahverdyan}, \citenamefont {Balian},\ and\ \citenamefont {Nieuwenhuizen}}]{E1}%
  \BibitemOpen
  \bibfield  {author} {\bibinfo {author} {\bibfnamefont {A.~E.}\ \bibnamefont {Allahverdyan}}, \bibinfo {author} {\bibfnamefont {R.}~\bibnamefont {Balian}}, \ and\ \bibinfo {author} {\bibfnamefont {Th.~M.}\ \bibnamefont {Nieuwenhuizen}},\ }\bibfield  {title} {\enquote {\bibinfo {title} {Maximal work extraction from finite quantum systems},}\ }\href {https://iopscience.iop.org/article/10.1209/epl/i2004-10101-2/meta} {\bibfield  {journal} {\bibinfo  {journal} {EPL}\ }\textbf {\bibinfo {volume} {67}},\ \bibinfo {pages} {565} (\bibinfo {year} {2004}{\natexlab{a}})}\BibitemShut {NoStop}%
\bibitem [{\citenamefont {Lenard}(1978)}]{P1}%
  \BibitemOpen
  \bibfield  {author} {\bibinfo {author} {\bibfnamefont {A.}~\bibnamefont {Lenard}},\ }\bibfield  {title} {\enquote {\bibinfo {title} {Thermodynamical proof of the gibbs formula for elementary quantum systems},}\ }\href {https://link.springer.com/article/10.1007/BF01011769} {\bibfield  {journal} {\bibinfo  {journal} {J. Stat. Phys.}\ }\textbf {\bibinfo {volume} {19}},\ \bibinfo {pages} {586} (\bibinfo {year} {1978})}\BibitemShut {NoStop}%
\bibitem [{\citenamefont {Pusz}\ and\ \citenamefont {Woronowicz}(1978)}]{P2}%
  \BibitemOpen
  \bibfield  {author} {\bibinfo {author} {\bibfnamefont {W.}~\bibnamefont {Pusz}}\ and\ \bibinfo {author} {\bibfnamefont {S.~L.}\ \bibnamefont {Woronowicz}},\ }\bibfield  {title} {\enquote {\bibinfo {title} {Passive states and kms states for general quantum systems},}\ }\href {https://link.springer.com/article/10.1007/BF01614224} {\bibfield  {journal} {\bibinfo  {journal} {Commun. Math. Phys.}\ }\textbf {\bibinfo {volume} {58}},\ \bibinfo {pages} {290} (\bibinfo {year} {1978})}\BibitemShut {NoStop}%
\bibitem [{\citenamefont {Perarnau-Llobet}\ \emph {et~al.}(2015)\citenamefont {Perarnau-Llobet}, \citenamefont {Hovhannisyan}, \citenamefont {Huber}, \citenamefont {Skrzypczyk}, \citenamefont {Tura},\ and\ \citenamefont {Ac\'{\i}n}}]{P3}%
  \BibitemOpen
  \bibfield  {author} {\bibinfo {author} {\bibfnamefont {M.}~\bibnamefont {Perarnau-Llobet}}, \bibinfo {author} {\bibfnamefont {K.~V.}\ \bibnamefont {Hovhannisyan}}, \bibinfo {author} {\bibfnamefont {M.}~\bibnamefont {Huber}}, \bibinfo {author} {\bibfnamefont {P.}~\bibnamefont {Skrzypczyk}}, \bibinfo {author} {\bibfnamefont {J.}~\bibnamefont {Tura}}, \ and\ \bibinfo {author} {\bibfnamefont {A.}~\bibnamefont {Ac\'{\i}n}},\ }\bibfield  {title} {\enquote {\bibinfo {title} {Most energetic passive states},}\ }\href {https://link.aps.org/doi/10.1103/PhysRevE.92.042147} {\bibfield  {journal} {\bibinfo  {journal} {Phys. Rev. E}\ }\textbf {\bibinfo {volume} {92}},\ \bibinfo {pages} {042147} (\bibinfo {year} {2015})}\BibitemShut {NoStop}%
\bibitem [{\citenamefont {Skrzypczyk}\ \emph {et~al.}(2015)\citenamefont {Skrzypczyk}, \citenamefont {Silva},\ and\ \citenamefont {Brunner}}]{P4}%
  \BibitemOpen
  \bibfield  {author} {\bibinfo {author} {\bibfnamefont {P.}~\bibnamefont {Skrzypczyk}}, \bibinfo {author} {\bibfnamefont {R.}~\bibnamefont {Silva}}, \ and\ \bibinfo {author} {\bibfnamefont {N.}~\bibnamefont {Brunner}},\ }\bibfield  {title} {\enquote {\bibinfo {title} {Passivity, complete passivity, and virtual temperatures},}\ }\href {https://link.aps.org/doi/10.1103/PhysRevE.91.052133} {\bibfield  {journal} {\bibinfo  {journal} {Phys. Rev. E}\ }\textbf {\bibinfo {volume} {91}},\ \bibinfo {pages} {052133} (\bibinfo {year} {2015})}\BibitemShut {NoStop}%
\bibitem [{\citenamefont {Brown}\ \emph {et~al.}(2016)\citenamefont {Brown}, \citenamefont {Friis},\ and\ \citenamefont {Huber}}]{P5}%
  \BibitemOpen
  \bibfield  {author} {\bibinfo {author} {\bibfnamefont {E.~G.}\ \bibnamefont {Brown}}, \bibinfo {author} {\bibfnamefont {N.}~\bibnamefont {Friis}}, \ and\ \bibinfo {author} {\bibfnamefont {M.}~\bibnamefont {Huber}},\ }\bibfield  {title} {\enquote {\bibinfo {title} {Passivity and practical work extraction using gaussian operations},}\ }\href {https://iopscience.iop.org/article/10.1088/1367-2630/18/11/113028} {\bibfield  {journal} {\bibinfo  {journal} {NJP}\ }\textbf {\bibinfo {volume} {18}},\ \bibinfo {pages} {113028} (\bibinfo {year} {2016})}\BibitemShut {NoStop}%
\bibitem [{\citenamefont {Sparaciari}\ \emph {et~al.}(2017)\citenamefont {Sparaciari}, \citenamefont {Jennings},\ and\ \citenamefont {Oppenheim}}]{P6}%
  \BibitemOpen
  \bibfield  {author} {\bibinfo {author} {\bibfnamefont {C.}~\bibnamefont {Sparaciari}}, \bibinfo {author} {\bibfnamefont {D.}~\bibnamefont {Jennings}}, \ and\ \bibinfo {author} {\bibfnamefont {J.}~\bibnamefont {Oppenheim}},\ }\bibfield  {title} {\enquote {\bibinfo {title} {Energetic instability of passive states in thermodynamics},}\ }\href {https://www.nature.com/articles/s41467-017-01505-4} {\bibfield  {journal} {\bibinfo  {journal} {Nat. Commun}\ }\textbf {\bibinfo {volume} {8}},\ \bibinfo {pages} {1895} (\bibinfo {year} {2017})}\BibitemShut {NoStop}%
\bibitem [{\citenamefont {Alhambra}\ \emph {et~al.}(2019{\natexlab{b}})\citenamefont {Alhambra}, \citenamefont {Styliaris}, \citenamefont {Rodr\'{\i}guez-Briones}, \citenamefont {Sikora},\ and\ \citenamefont {Mart\'{\i}n-Mart\'{\i}nez}}]{P7}%
  \BibitemOpen
  \bibfield  {author} {\bibinfo {author} {\bibfnamefont {\'A.~M.}\ \bibnamefont {Alhambra}}, \bibinfo {author} {\bibfnamefont {G.}~\bibnamefont {Styliaris}}, \bibinfo {author} {\bibfnamefont {N.~A.}\ \bibnamefont {Rodr\'{\i}guez-Briones}}, \bibinfo {author} {\bibfnamefont {J.}~\bibnamefont {Sikora}}, \ and\ \bibinfo {author} {\bibfnamefont {E.}~\bibnamefont {Mart\'{\i}n-Mart\'{\i}nez}},\ }\bibfield  {title} {\enquote {\bibinfo {title} {Fundamental limitations to local energy extraction in quantum systems},}\ }\href {https://link.aps.org/doi/10.1103/PhysRevLett.123.190601} {\bibfield  {journal} {\bibinfo  {journal} {Phys. Rev. Lett.}\ }\textbf {\bibinfo {volume} {123}},\ \bibinfo {pages} {190601} (\bibinfo {year} {2019}{\natexlab{b}})}\BibitemShut {NoStop}%
\bibitem [{\citenamefont {Maillette~de Buy~Wenniger}\ \emph {et~al.}(2023)\citenamefont {Maillette~de Buy~Wenniger}, \citenamefont {Thomas}, \citenamefont {Maffei}, \citenamefont {Wein}, \citenamefont {Pont}, \citenamefont {Belabas}, \citenamefont {Prasad}, \citenamefont {Harouri}, \citenamefont {Lema\^{\i}tre}, \citenamefont {Sagnes}, \citenamefont {Somaschi}, \citenamefont {Auff\`eves},\ and\ \citenamefont {Senellart}}]{qd}%
  \BibitemOpen
  \bibfield  {author} {\bibinfo {author} {\bibfnamefont {I.}~\bibnamefont {Maillette~de Buy~Wenniger}}, \bibinfo {author} {\bibfnamefont {S.~E.}\ \bibnamefont {Thomas}}, \bibinfo {author} {\bibfnamefont {M.}~\bibnamefont {Maffei}}, \bibinfo {author} {\bibfnamefont {S.~C.}\ \bibnamefont {Wein}}, \bibinfo {author} {\bibfnamefont {M.}~\bibnamefont {Pont}}, \bibinfo {author} {\bibfnamefont {N.}~\bibnamefont {Belabas}}, \bibinfo {author} {\bibfnamefont {S.}~\bibnamefont {Prasad}}, \bibinfo {author} {\bibfnamefont {A.}~\bibnamefont {Harouri}}, \bibinfo {author} {\bibfnamefont {A.}~\bibnamefont {Lema\^{\i}tre}}, \bibinfo {author} {\bibfnamefont {I.}~\bibnamefont {Sagnes}}, \bibinfo {author} {\bibfnamefont {N.}~\bibnamefont {Somaschi}}, \bibinfo {author} {\bibfnamefont {A.}~\bibnamefont {Auff\`eves}}, \ and\ \bibinfo {author} {\bibfnamefont {P.}~\bibnamefont {Senellart}},\ }\bibfield  {title} {\enquote {\bibinfo {title} {Experimental analysis of energy transfers between a quantum emitter and light fields},}\ }\href
  {https://link.aps.org/doi/10.1103/PhysRevLett.131.260401} {\bibfield  {journal} {\bibinfo  {journal} {Phys. Rev. Lett.}\ }\textbf {\bibinfo {volume} {131}},\ \bibinfo {pages} {260401} (\bibinfo {year} {2023})}\BibitemShut {NoStop}%
\bibitem [{\citenamefont {Le}\ \emph {et~al.}(2018{\natexlab{b}})\citenamefont {Le}, \citenamefont {Levinsen}, \citenamefont {Modi}, \citenamefont {Parish},\ and\ \citenamefont {Pollock}}]{ma}%
  \BibitemOpen
  \bibfield  {author} {\bibinfo {author} {\bibfnamefont {Thao~P.}\ \bibnamefont {Le}}, \bibinfo {author} {\bibfnamefont {Jesper}\ \bibnamefont {Levinsen}}, \bibinfo {author} {\bibfnamefont {Kavan}\ \bibnamefont {Modi}}, \bibinfo {author} {\bibfnamefont {Meera~M.}\ \bibnamefont {Parish}}, \ and\ \bibinfo {author} {\bibfnamefont {Felix~A.}\ \bibnamefont {Pollock}},\ }\bibfield  {title} {\enquote {\bibinfo {title} {Spin-chain model of a many-body quantum battery},}\ }\href {\doibase 10.1103/PhysRevA.97.022106} {\bibfield  {journal} {\bibinfo  {journal} {Phys. Rev. A}\ }\textbf {\bibinfo {volume} {97}},\ \bibinfo {pages} {022106} (\bibinfo {year} {2018}{\natexlab{b}})}\BibitemShut {NoStop}%
\bibitem [{\citenamefont {Liu}\ and\ \citenamefont {Hanna}(2018)}]{OM}%
  \BibitemOpen
  \bibfield  {author} {\bibinfo {author} {\bibfnamefont {J.}~\bibnamefont {Liu}}\ and\ \bibinfo {author} {\bibfnamefont {G.}~\bibnamefont {Hanna}},\ }\bibfield  {title} {\enquote {\bibinfo {title} {Efficient and deterministic propagation of mixed quantum-classical liouville dynamics},}\ }\href {https://pubs.acs.org/doi/10.1021/acs.jpclett.8b01619} {\bibfield  {journal} {\bibinfo  {journal} {JPCL}\ }\textbf {\bibinfo {volume} {9}},\ \bibinfo {pages} {3933} (\bibinfo {year} {2018})}\BibitemShut {NoStop}%
\bibitem [{\citenamefont {Quach}\ \emph {et~al.}(2022)\citenamefont {Quach}, \citenamefont {McGhee}, \citenamefont {Ganzer}, \citenamefont {Rouse}, \citenamefont {Lovett}, \citenamefont {Gauger}, \citenamefont {Keeling}, \citenamefont {Cerullo}, \citenamefont {Lidzey},\ and\ \citenamefont {Virgili}}]{OM2}%
  \BibitemOpen
  \bibfield  {author} {\bibinfo {author} {\bibfnamefont {J.~Q.}\ \bibnamefont {Quach}}, \bibinfo {author} {\bibfnamefont {K.~E.}\ \bibnamefont {McGhee}}, \bibinfo {author} {\bibfnamefont {L.}~\bibnamefont {Ganzer}}, \bibinfo {author} {\bibfnamefont {D.~M.}\ \bibnamefont {Rouse}}, \bibinfo {author} {\bibfnamefont {B.~W.}\ \bibnamefont {Lovett}}, \bibinfo {author} {\bibfnamefont {E.~M.}\ \bibnamefont {Gauger}}, \bibinfo {author} {\bibfnamefont {J.}~\bibnamefont {Keeling}}, \bibinfo {author} {\bibfnamefont {G.}~\bibnamefont {Cerullo}}, \bibinfo {author} {\bibfnamefont {D.~G.}\ \bibnamefont {Lidzey}}, \ and\ \bibinfo {author} {\bibfnamefont {T.}~\bibnamefont {Virgili}},\ }\bibfield  {title} {\enquote {\bibinfo {title} {Superabsorption in an organic microcavity: Toward a quantum battery},}\ }\href {https://www.science.org/doi/10.1126/sciadv.abk3160} {\bibfield  {journal} {\bibinfo  {journal} {Sci. Adv.}\ }\textbf {\bibinfo {volume} {8}},\ \bibinfo {pages} {eabk3160} (\bibinfo {year} {2022})}\BibitemShut {NoStop}%
\bibitem [{\citenamefont {Santos}\ \emph {et~al.}(2019)\citenamefont {Santos}, \citenamefont {Cakmak},\ and\ \citenamefont {Zinner}}]{Scon}%
  \BibitemOpen
  \bibfield  {author} {\bibinfo {author} {\bibfnamefont {A.~C.}\ \bibnamefont {Santos}}, \bibinfo {author} {\bibfnamefont {S.}~\bibnamefont {Cakmak}, \bibfnamefont {B.~Campbell}}, \ and\ \bibinfo {author} {\bibfnamefont {N.~T.}\ \bibnamefont {Zinner}},\ }\bibfield  {title} {\enquote {\bibinfo {title} {Stable adiabatic quantum batteries},}\ }\href {https://link.aps.org/doi/10.1103/PhysRevE.100.032107} {\bibfield  {journal} {\bibinfo  {journal} {Phys. Rev. E}\ }\textbf {\bibinfo {volume} {100}},\ \bibinfo {pages} {032107} (\bibinfo {year} {2019})}\BibitemShut {NoStop}%
\bibitem [{\citenamefont {Friis}\ and\ \citenamefont {Huber}(2018)}]{Pc}%
  \BibitemOpen
  \bibfield  {author} {\bibinfo {author} {\bibfnamefont {N.}~\bibnamefont {Friis}}\ and\ \bibinfo {author} {\bibfnamefont {M.}~\bibnamefont {Huber}},\ }\bibfield  {title} {\enquote {\bibinfo {title} {Precision and work fluctuations in gaussian battery charging},}\ }\href {https://doi.org/10.22331/q-2018-04-23-61} {\bibfield  {journal} {\bibinfo  {journal} {Quantum}\ }\textbf {\bibinfo {volume} {2}},\ \bibinfo {pages} {61} (\bibinfo {year} {2018})}\BibitemShut {NoStop}%
\bibitem [{\citenamefont {Hu}\ \emph {et~al.}(2022)\citenamefont {Hu}, \citenamefont {Qiu}, \citenamefont {Souza}, \citenamefont {Yuan}, \citenamefont {Zhou}, \citenamefont {Zhang}, \citenamefont {Chu}, \citenamefont {Pan},\ and\ \citenamefont {\textit{et.al}}}]{Ex1}%
  \BibitemOpen
  \bibfield  {author} {\bibinfo {author} {\bibfnamefont {C.-K.}\ \bibnamefont {Hu}}, \bibinfo {author} {\bibfnamefont {J.}~\bibnamefont {Qiu}}, \bibinfo {author} {\bibfnamefont {P.~J.}\ \bibnamefont {Souza}}, \bibinfo {author} {\bibfnamefont {J.}~\bibnamefont {Yuan}}, \bibinfo {author} {\bibfnamefont {Y.}~\bibnamefont {Zhou}}, \bibinfo {author} {\bibfnamefont {L.}~\bibnamefont {Zhang}}, \bibinfo {author} {\bibfnamefont {J.}~\bibnamefont {Chu}}, \bibinfo {author} {\bibfnamefont {X.}~\bibnamefont {Pan}}, \ and\ \bibinfo {author} {\bibnamefont {\textit{et.al}}},\ }\bibfield  {title} {\enquote {\bibinfo {title} {Optimal charging of a superconducting quantum battery},}\ }\href {https://iopscience.iop.org/article/10.1088/2058-9565/ac8444/meta} {\bibfield  {journal} {\bibinfo  {journal} {QST}\ }\textbf {\bibinfo {volume} {7}},\ \bibinfo {pages} {045018} (\bibinfo {year} {2022})}\BibitemShut {NoStop}%
\bibitem [{\citenamefont {Joshi}\ and\ \citenamefont {Mahesh}(2022)}]{Ex2}%
  \BibitemOpen
  \bibfield  {author} {\bibinfo {author} {\bibfnamefont {J.}~\bibnamefont {Joshi}}\ and\ \bibinfo {author} {\bibfnamefont {T.~S.}\ \bibnamefont {Mahesh}},\ }\bibfield  {title} {\enquote {\bibinfo {title} {Experimental investigation of a quantum battery using star-topology nmr spin systems},}\ }\href {https://link.aps.org/doi/10.1103/PhysRevA.106.042601} {\bibfield  {journal} {\bibinfo  {journal} {Phys. Rev. A}\ }\textbf {\bibinfo {volume} {106}},\ \bibinfo {pages} {042601} (\bibinfo {year} {2022})}\BibitemShut {NoStop}%
\bibitem [{\citenamefont {Quach}\ and\ \citenamefont {Munro}(2020)}]{Ex3}%
  \BibitemOpen
  \bibfield  {author} {\bibinfo {author} {\bibfnamefont {J.~Q.}\ \bibnamefont {Quach}}\ and\ \bibinfo {author} {\bibfnamefont {W.~J.}\ \bibnamefont {Munro}},\ }\bibfield  {title} {\enquote {\bibinfo {title} {Using dark states to charge and stabilize open quantum batteries},}\ }\href {https://link.aps.org/doi/10.1103/PhysRevApplied.14.024092} {\bibfield  {journal} {\bibinfo  {journal} {Phys. Rev. Appl.}\ }\textbf {\bibinfo {volume} {14}},\ \bibinfo {pages} {024092} (\bibinfo {year} {2020})}\BibitemShut {NoStop}%
\bibitem [{\citenamefont {Allahverdyan}\ \emph {et~al.}(2004{\natexlab{b}})\citenamefont {Allahverdyan}, \citenamefont {Balian},\ and\ \citenamefont {Nieuwenhuizen}}]{Er2}%
  \BibitemOpen
  \bibfield  {author} {\bibinfo {author} {\bibfnamefont {A.~E.}\ \bibnamefont {Allahverdyan}}, \bibinfo {author} {\bibfnamefont {R.}~\bibnamefont {Balian}}, \ and\ \bibinfo {author} {\bibfnamefont {Th.~M.}\ \bibnamefont {Nieuwenhuizen}},\ }\bibfield  {title} {\enquote {\bibinfo {title} {Maximal work extraction from finite quantum systems},}\ }\href {https://iopscience.iop.org/article/10.1209/epl/i2004-10101-2} {\bibfield  {journal} {\bibinfo  {journal} {EPL}\ }\textbf {\bibinfo {volume} {67}},\ \bibinfo {pages} {565} (\bibinfo {year} {2004}{\natexlab{b}})}\BibitemShut {NoStop}%
\bibitem [{\citenamefont {Bhattacharjee}\ and\ \citenamefont {Dutta}(2021{\natexlab{b}})}]{Er3}%
  \BibitemOpen
  \bibfield  {author} {\bibinfo {author} {\bibfnamefont {S.}~\bibnamefont {Bhattacharjee}}\ and\ \bibinfo {author} {\bibfnamefont {A.}~\bibnamefont {Dutta}},\ }\bibfield  {title} {\enquote {\bibinfo {title} {Quantum thermal machines and batteries},}\ }\href {https://link.springer.com/article/10.1140/epjb/s10051-021-00235-3} {\bibfield  {journal} {\bibinfo  {journal} {EPJ B}\ }\textbf {\bibinfo {volume} {94}},\ \bibinfo {pages} {42} (\bibinfo {year} {2021}{\natexlab{b}})}\BibitemShut {NoStop}%
\bibitem [{\citenamefont {Aberg}(2006)}]{Co1}%
  \BibitemOpen
  \bibfield  {author} {\bibinfo {author} {\bibfnamefont {J.}~\bibnamefont {Aberg}},\ }\bibfield  {title} {\enquote {\bibinfo {title} {Quantifying superposition},}\ }\href {https://doi.org/10.48550/arXiv.quant-ph/0612146} {\bibfield  {journal} {\bibinfo  {journal} {arXiv preprint quant-ph/0612146}\ } (\bibinfo {year} {2006})}\BibitemShut {NoStop}%
\bibitem [{\citenamefont {Baumgratz}\ \emph {et~al.}(2014)\citenamefont {Baumgratz}, \citenamefont {Cramer},\ and\ \citenamefont {Plenio}}]{Co2}%
  \BibitemOpen
  \bibfield  {author} {\bibinfo {author} {\bibfnamefont {T.}~\bibnamefont {Baumgratz}}, \bibinfo {author} {\bibfnamefont {M.}~\bibnamefont {Cramer}}, \ and\ \bibinfo {author} {\bibfnamefont {M.~B.}\ \bibnamefont {Plenio}},\ }\bibfield  {title} {\enquote {\bibinfo {title} {Quantifying coherence},}\ }\href {\doibase 10.1103/PhysRevLett.113.140401} {\bibfield  {journal} {\bibinfo  {journal} {Phys. Rev. Lett.}\ }\textbf {\bibinfo {volume} {113}},\ \bibinfo {pages} {140401} (\bibinfo {year} {2014})}\BibitemShut {NoStop}%
\bibitem [{\citenamefont {Levi}\ and\ \citenamefont {Mintert}(2014)}]{Co3}%
  \BibitemOpen
  \bibfield  {author} {\bibinfo {author} {\bibfnamefont {F.}~\bibnamefont {Levi}}\ and\ \bibinfo {author} {\bibfnamefont {F.}~\bibnamefont {Mintert}},\ }\bibfield  {title} {\enquote {\bibinfo {title} {A quantitative theory of coherent delocalization},}\ }\href@noop {} {\bibfield  {journal} {\bibinfo  {journal} {New J. Phys.}\ }\textbf {\bibinfo {volume} {16}},\ \bibinfo {pages} {033007} (\bibinfo {year} {2014})}\BibitemShut {NoStop}%
\bibitem [{\citenamefont {Winter}\ and\ \citenamefont {Yang}(2016)}]{Co4}%
  \BibitemOpen
  \bibfield  {author} {\bibinfo {author} {\bibfnamefont {A.}~\bibnamefont {Winter}}\ and\ \bibinfo {author} {\bibfnamefont {D.}~\bibnamefont {Yang}},\ }\bibfield  {title} {\enquote {\bibinfo {title} {Operational resource theory of coherence},}\ }\href {https://link.aps.org/doi/10.1103/PhysRevLett.116.120404} {\bibfield  {journal} {\bibinfo  {journal} {Phys. Rev. Lett.}\ }\textbf {\bibinfo {volume} {116}},\ \bibinfo {pages} {120404} (\bibinfo {year} {2016})}\BibitemShut {NoStop}%
\bibitem [{\citenamefont {Streltsov}\ \emph {et~al.}(2017)\citenamefont {Streltsov}, \citenamefont {Adesso},\ and\ \citenamefont {Plenio}}]{Co5}%
  \BibitemOpen
  \bibfield  {author} {\bibinfo {author} {\bibfnamefont {A.}~\bibnamefont {Streltsov}}, \bibinfo {author} {\bibfnamefont {G.}~\bibnamefont {Adesso}}, \ and\ \bibinfo {author} {\bibfnamefont {M.~B.}\ \bibnamefont {Plenio}},\ }\bibfield  {title} {\enquote {\bibinfo {title} {Colloquium: Quantum coherence as a resource},}\ }\href {https://link.aps.org/doi/10.1103/RevModPhys.89.041003} {\bibfield  {journal} {\bibinfo  {journal} {Rev. Mod. Phys.}\ }\textbf {\bibinfo {volume} {89}},\ \bibinfo {pages} {041003} (\bibinfo {year} {2017})}\BibitemShut {NoStop}%
\end{thebibliography}%
\end{document}